\documentclass[pra,aps,showpacs]{revtex4}

\usepackage{amssymb}
\usepackage{amsmath}
\usepackage{array}
\usepackage{graphicx}

\def\c{\cdot}

\def\Gr{{\mathcal G}}
\def\eps{\varepsilon}
\def\adj{^\dagger}
\def\ktau{(\vec{k},\tau)}
\def\komega{(\vec{k},i\omega{_n})}

\renewcommand\vec[1]{{\mathbf #1}}
\def\mat[#1][#2]{{\hspace{.3ex}{\frac{}{}}^{#1}\underline{\underline{#2}}}}
\def\Gr{{\mathcal G}}
\def\gmat[#1][#2]{\hspace{.3ex}{\frac{}{}}^{#1} #2}
\def\vv[#1][#2][#3]{\hspace{.3ex}{\frac{}{}}_{#1}^{#2} \underline{#3}}
\def\kink{e_{\vec{k}}}

\def\Im{ \mathfrak{Im} }
\def\Re{ \mathfrak{Re}}

\newcommand\epss[1]{ {{\mat[#1][\eps]}^\star} }
\def\nul{_{\raisebox{1ex}{\scriptsize{(0)}}}}
\def\ws{\widehat \sigma}

\begin{document}
\title{Energies and damping rates of elementary excitations in spin-1
  {B}ose-{E}instein condensed gases}
\author{Gergely Szirmai}
\affiliation{Department of Physics of Complex Systems, Roland E{\"o}tv{\"o}s
University, P{\'a}zm{\'a}ny P{\'e}ter s{\'e}t{\'a}ny 1/A, Budapest, H-1117}
\affiliation{Research Group for Statistical Physics of the Hungarian Academy
  of Sciences, P{\'a}zm{\'a}ny P{\'e}ter S{\'e}t{\'a}ny 1/A, Budapest, H-1117}
\author{P{\'e}ter Sz{\'e}pfalusy}
\affiliation{Department of Physics of Complex Systems, Roland E{\"o}tv{\"o}s
University, P{\'a}zm{\'a}ny P{\'e}ter s{\'e}t{\'a}ny 1/A, Budapest, H-1117}
\affiliation{Research Institute for Solid State Physics and Optics of the
Hungarian Academy of Sciences, Budapest, P.O.Box 49, H-1525}
\author{Kriszti{\'a}n Kis-Szab{\'o}}
\affiliation{Department of Physics of Complex Systems, Roland E{\"o}tv{\"o}s
University, P{\'a}zm{\'a}ny P{\'e}ter s{\'e}t{\'a}ny 1/A, Budapest, H-1117}
\date\today
\begin{abstract}
  Finite temperature Green's function technique is used to calculate
  the energies and damping rates of elementary excitations of the
  homogeneous, dilute, spin-1 Bose gases below the Bose-Einstein
  condensation temperature both in the density and spin channels. For
  this purpose the self-consistent dynamical Hartree-Fock model is
  formulated, which takes into account the direct and exchange
  processes on equal footing by summing up certain classes of Feynman
  diagrams. The model is shown to fulfil the Goldstone theorem and to
  exhibit the hybridization of one-particle and collective excitations
  correctly. The results are applied to the gases of
  $^{23}\mathrm{Na}$ and $^{87}\mathrm{Rb}$ atoms.
\end{abstract}

\pacs{03.75.Mn, 03.75.Kk, 03.75Hh}

\maketitle

\section{Introduction}

The experimental realization of {B}ose-{E}instein condensation with
ultracold, dilute, alkali gases in magnetic traps \cite{Aea2,Dea} made
it possible, among other things, to meaningfully compare the results
of experiment and theory. The reason lies in the diluteness of the gas
sample, so the interaction between the atoms of the gas can be modeled
by a week pseudopotential, in most of the cases, which can be
considered as perturbation in calculations. With the development of
optical traps it became possible to confine not just one or two of the
magnetic states $\left|f,m\right>$ of the atom (the weak-field seeking
states), but all of them for a given hyperfine spin $f$ giving rise to
a so called spinor condensate
\cite{SKea,Sea1,Sea2,Miesea,Stamper-Kurn2001a}. Because of these
internal degrees of freedom such systems can have much wider dynamical
properties than those confined in magnetic traps. The theoretical
study of spinor Bose condensed systems was initiated by \cite{Ho2,OM}.

There are two different phases of spin-1 systems in the presence of a
Bose condensate distinguished according to the total magnetic moment
of the system \cite{Ho2,OM}.  Namely, the \emph{ferromagnetic phase},
which is characterized by the existence of a macroscopic magnetic
moment and the \emph{polar phase} where the system has no macroscopic
magnetic moment at all. In the absence of an external magnetic field
the type of the phase of the spin-1 Bose system depends only on the
interaction potential of the atoms, which can be accurately
characterized by s-wave scattering lengths $a_0$ and $a_2$ responsible
for scatterings in the total hyperfine spin channel zero and two,
respectively \cite{Ho2}. If $a_2 > a_0$ the system prefers the polar
phase, while in the case of $a_2 < a_0$ the system is ferromagnetic
\cite{Ho2}. For both cases there exist species of alkali Bose atoms.
For example the $f=1$ hyperfine spin state of $^{23}\mathrm{Na}$
realizes the polar phase, while that of $^{87}\mathrm{Rb}$ the
ferromagnetic phase. Since in the absence of a magnetic field the
phase of the system does not depend on experimentally tunable
parameters, one can use the word 'case' instead of 'phase'.

The elementary excitations of dilute Bose gases have been studied for
a long time \cite{Bogolj,Belia}, since these gases represent the
simplest prototypes of symmetry breaking quantum systems. The zero
temperature elementary excitations of these systems are described
reasonably well by the Bogoliubov Hamiltonian \cite{Bogolj}.  It is
not straightforward to generalize Bogoliubov's theory to finite
temperatures in a consistent way
\cite{SG,Griffin,Griffin2,Popov1,SzK,FRSzG,SzSz1,SzSz2}.  The simplest
finite temperature approach is the self-consistent Hartree model, also
known as the random phase approximation \cite{SzK,SzSz2,Griffin}.

The inclusion of exchange processes is not trivial in a consistent way
using the finite temperature Green's function technique as shown in
\cite{FRSzG} for a scalar Bose-gas. (See for a formulation of the
Hartree-Fock theory for such systems giving the density response
functions \cite{MT1}) For instance self-consistency has proven to be
essential for developing a theory satisfying general requirements. As
a matter of fact applying unperturbed (i.e. free particle) propagators
as internal line contributions leads to an instability for long
wavelength fluctuations in case of repulsive interactions.  (Note,
however, that for attractive interactions such model has been used to
describe the collapse of the gas phase \cite{MB1}.)

In the present paper we generalize the model presented in \cite{FRSzG}
with systems with scalar condensate to the spin-1 Bose gas. Throughout
the paper we will work in the orderd phase.  We start from the
self-consistent Hartree-Fock approximation for the equation of state,
whose iteration we reveal which contributions are included in the
self-consistent procedure. On this basis the self-energies of various
one-particle Green's functions can be given in harmony with the
equation of state. Furthermore the obtained self-energies fulfill the
Goldstone theorem, which is the consequence of the breaking of the
gauge and spin rotation symmetries. This ensures gapless excitation
spectra.  Having determined the one-particle propagators, the
expressions of various correlation funcions will be given. The
requirement of the coincidence of the one-particle and certain
collective excitations is satisfied and the model resulted fits into
the dielectric formalism. (For a detailed discussion of the dielectric
formalism for spin-1 systems see \cite{SzSz2} and for a general review
\cite{Griffin}.) One has to point out that the model developed can be
regarded as an improved mean-field one and as such looses its validity
near the phase transition point.

Since the present work deals with homogeneous systems, the results are
applicable to trapped gases only when a local density approximation is
relevant or when the quantities in question are not sensitive to the
finiteness of the system. In analogy with what proved to be the
situation in the case of gases with a scalar condensate (see
\cite{FRSzG,Rea1} and references therein) one can hope that the
damping of the elementary excitations belongs to the latter category.
Concerning the local density approach it remains to be seen whether
such experiment which provided the frequency of local excitations
\cite{Aea} can be carried out for spin-1 gases. Anyhow, we evaluate
the characteristics of the excitations for parameter values which
might be relevant to applications. Finally we note that a
generalization of the model to finite systems is straightforward like
for systems with a scalar condensate \cite {Rea1,BSz,Rea2}.

We hope to carry out the calculation directly to trapped systems in
the future. At this point one has to emphasize again the
self-consistent nature of the model which is very important for
applications to confined gases.

The outline of the paper is as follows. In Sec. \ref{sec:form}. the
formulation of the Hartree-Fock approximation is given for the spin-1
Bose gas. It includes the specification of the Hamiltonian, the
dealing with the symmetry breaking by a canonical transformation, the
equation of state, the definitions and properties the Green's
functions and of various correlation functions. In this section we
specify what building blocks are to be used for the determination of
the various quantities such that
\begin{itemize}
\item being consistent with the implicit summation occurring in the
  Hartree-Fock propagator due to the self-consistency condition,
\item fullfilling the hybridization  property.
\end{itemize}
In Sec. \ref{sec:appl}. the Hartree-Fock model is applied to a
ferromagnetic system, the gas of $^{87}\mathrm{Rb}$ atoms [Sec.
\ref{ssec:ferr}], and to the $^{23}\mathrm{Na}$ gas, which realizes
the polar phase [Sec. \ref{ssec:polar}]. Sec. \ref{sec:conc}. is
devoted for further discussion also focusing on an overview of the
physical properties of the various collective modes considered in
Sec. \ref{sec:appl}. Some details are relegated to the Appendix.

\section{Formulation}
\label{sec:form}

We consider a system of spin-1 bosons in a unit volume box with
periodic boundary conditions. For such a system the plausible basis
set of one particle states to build the Fock-space from are plane wave
states. The creation and destruction operators are $a\adj_r(\vec{k})$
and $a_r(\vec{k})$, which create and annihilate a one-particle state
with momentum $\vec{k}$ and spin projection $r$, respectively.
Throughout the paper Roman indices are used for distinguishing between
spin components of a quantity. These indices will always take values
from the set $\{+,0,-\}$ referring to the eigenvalue of $F_z$. Unless
otherwise stated we understand automatic summation over repeated Roman
indices. For our purposes, while the temperature of the system is low,
and the gas is very dilute, the system can be well described by the
following low energy, effective, grand-canonical Hamiltonian
\cite{Ho2}:
\begin{equation}
  \label{eq:ham}
  {\mathcal H}=\sum_{\vec{k}}(e_{\vec{k}}-\mu)a_r^\dagger(\vec{k})
  a_r(\vec{k})+\frac{1}{2}\sum_{\vec{k}_1+\vec{k}_2=\vec{k}_3
    +\vec{k}_4}a^\dagger_{r'}(\vec{k}_1)a^\dagger_r(\vec{k}_2)V^{r's'}_{rs}
  a_s(\vec{k}_3)a_{s'}(\vec{k}_4),
\end{equation}
where $e_{\vec{k}}=\hslash^2 k^2/2M$ is the kinetic energy of a
particle with mass $M$, and $\mu$ is the chemical potential. We
restrict ourselves to the Hartree-Fock approximation, and for such
purposes the interatomic potential in Eq. \eqref{eq:ham} can be
modeled by the momentum independent pseudopotential \cite{BS1,SG}:
\begin{equation}
  \label{eq:pseudopot}
    V^{r's'}_{rs}=c_n\delta_{rs}\delta_{r's'}+c_s(\vec{F})_{rs}
    (\vec{F})_{r's'},
\end{equation}
with parameters:
\begin{subequations}
  \label{eqs:pspars}
  \begin{align}
    c_n&=\frac{4\pi\hslash^2}{M}\frac{a_0+2a_2}{3},\\
    c_s&=\frac{4\pi\hslash^2}{M}\frac{a_2-a_0}{3}.
  \end{align}
\end{subequations}
The parameters $a_0$ and $a_2$ are the scattering length in the total
hyperfine spin channel zero and two, respectively. Note, that $c_s>0$
for the polar case and $c_s<0$ for the ferromagnetic one. In our
representation the one-particle spin operators are
\begin{equation}
  \label{eq:spinops}
  F_x=\frac{1}{\sqrt{2}}\left [ \begin{array}{c c c}
      0&1&0\\
      1&0&1\\
      0&1&0
    \end{array}\right ], \qquad
  F_y=\frac{1}{\sqrt{2}}\left [ \begin{array}{c c c}
      0&-i&0\\
      i&0&-i\\
      0&i&0
    \end{array}\right ], \qquad
  F_z=\left [ \begin{array}{c c c}
      1&0&0\\
      0&0&0\\
      0&0&-1
    \end{array}\right ].\qquad\qquad\qquad\quad
\end{equation}

Let us denote the temperature with $T_0$, where the symmetric
(uncondensed or high temperature) phase becomes unstable. To treat the
system in the Bose condensed phase one can introduce the following new
set of operators with a canonical transformation:
\begin{subequations}\label{eqs:cantr}
  \begin{align}
    b_r(\vec{k})=a_r(\vec{k})-\delta_{\vec{k},0}
    \sqrt{N_0}\zeta_r,\label{eq:cantr1}\\
    b_r^\dagger(\vec{k})=a_r^\dagger(\vec{k})-\delta_{\vec{k},0}
    \sqrt{N_0}\zeta_r\adj.\label{eq:cantr2}
  \end{align}
\end{subequations}
The $N_0$ and $\zeta_r$ parameters appearing in the transformation
\eqref{eqs:cantr} describe the number and spinor of particles in the
condensate, respectively. The spinor is normalized to unity. The
spinor describing the polar state can be taken as $\zeta=(0, 1, 0)^T$,
while the ferromagnetic spinor as $\zeta=(1, 0, 0)^T$. Here and in the
following, the '$T$' in the superscript means transposition.  The
equation of state consists of two parts. The first one is the
condition
\begin{subequations}
  \label{eqs:eqstates}
  \begin{equation}
    \label{eq:vanelexc}
    \left< b_r (\vec{k}) \right> = \left< b_r\adj (\vec{k}) \right> = 0,
  \end{equation}
  which provides an equation among $N_0$ and the chemical potential
  (see e.g. Ref.  \cite{SzSz2}). The other part is the expression of
  the particle number. In our case
  \begin{equation}
    \label{eq:eqstate1}
    N=\sum_{\vec{k}}\left<a_r\adj(\vec{k})a_r(\vec{k})\right>=N_0(T,\mu)+
    \sum_{\vec{k}}\left<b_r\adj(\vec{k})b_r(\vec{k})\right>,
  \end{equation}
\end{subequations}
where we have used Eqs. \eqref{eqs:cantr} to arrive to the right hand
side of the above equation.

\subsection{Green's functions}

To calculate the grand-canonical averages in the equation of state
\eqref{eqs:eqstates} we use the method of finite temperature Green's
function theory. The Green's functions of the system are defined by:
\begin{equation}
  \label{eq:grdef}
      \Gr^{rs}_{\gamma\delta}\ktau:=-\Big<T_\tau\big[b^\gamma_r\ktau
    b^{\delta\adj}_s(\vec{k},0)\big]\Big>,
\end{equation}
where $\tau$ stands for the imaginary time and $T_\tau$ is the $\tau$
ordering operator. The Greek indices are introduced for abbreviation,
with $b_r^1(\vec{k})=b_r(\vec{k})$ and
$b_r^{-1}(\vec{k})=b_r\adj(-\vec{k})$. Since they show the direction
of propagation of the Green's function, we also will refer to them as
direction indices. The convention of automatic summation is also
understood for the Greek indices unless stated otherwise.

These functions are periodic in $\tau$ with period $\beta\hslash$. It
is more convenient to use the Fourier series instead of the imaginary
time dependent functions:
\begin{equation}
  \label{eq:matsrep}
  \Gr^{rs}_{\gamma\delta}\komega=\frac{1}{\beta\hslash}
  \int_0^{\beta\hslash}d\tau e^{i\omega_n\tau}\Gr^{rs}_{\gamma\delta}\ktau,
\end{equation}
with $\omega_n=2n\pi/\beta\hslash$ the Bose discrete Matsubara
frequency \cite{FW}.

The perturbation series of the Green's functions permit a simple
partial summation. The Green's functions satisfy the generalized
Dyson-Beliaev equations \cite{SzSz2,Griffin,FW}:
\begin{equation}
  \label{eq:dyson}
  \Gr^{rs}_{\gamma\delta}\komega=\Gr^{rs}_{(0)\gamma\delta}\komega+\Gr^{rr'}
  _{(0)\gamma\rho}\komega\Sigma^{r's'}_{\rho\sigma}\komega\Gr^{s's}_{\sigma
    \delta}\komega,
\end{equation}
where
\begin{equation}
  \label{eq:freeprop}
  \Gr^{rs}_{(0)\gamma\delta}\komega=\frac{\delta_{rs}\delta_{\gamma\delta}}
  {i\omega_n-\hslash^{-1}\kink}
\end{equation}
is the free propagator, and $\Sigma^{r's'}_{\rho\sigma}\komega$ is the
self-energy, i.e. the sum of the contribution of the one particle
irreducible Feynman diagrams, which connect to two external lines, one
from the left, with indices $r',\rho$, momentum $\vec{k}$ and
frequency $\omega_n$, the other from the right, with indices
$s',\sigma$, momentum $k$ and frequency $\omega_n$ (we will call the one
particle irreducible graphs simply as irreducible graphs). Note, that
the chemical potential is included in the self energy instead of the
free-proagator (see, e.g., Ref. \cite{SzSz2}). Both the Green's
functions and self-energies obey the following symmetry properties
\cite{SzSz2}:
\begin{equation}
  \Gr^{rs}_{\gamma\delta}\komega=\Gr^{sr}_{\delta\gamma}\komega
  =\Gr^{sr}_{-\delta,-\gamma}(\vec{k},-i\omega_n),\label{eq:symeq1}
\end{equation}
where $\Sigma$ can also stand instead of $\Gr$ in Eq.
\eqref{eq:symeq1}. Since spin and space variables are not coupled, the
Green's functions and self-energies depend only on the modulus of the
momentum.

According to the usual procedure elementary excitations can be
obtained. Continuing the Green's functions in their frequency variable
from the imaginary Matsubara frequencies to the complex plane, requiring
to be 
 analytical on the upper-half plane, one arrives at the retarded
Green's functions, the poles of which (in the lower half-plane) give
the various one-particle elementary excitations of the system.

\subsection{The equation of state}

Now let us return to the problem of determining the equation of state
\eqref{eqs:eqstates} of the symmetry breaking dilute spin-1 Bose gas
in the Hartree-Fock approximation. Indeed, the equation of state in
the Hartree-Fock approximation is the system of nonlinear equations
resulting from the self-consistency condition of the Hartree-Fock
propagators along with Eqs. \eqref{eqs:eqstates}.

\begin{figure}[!ht]
  \centering
  \includegraphics*[30mm,236mm][190mm,278mm]{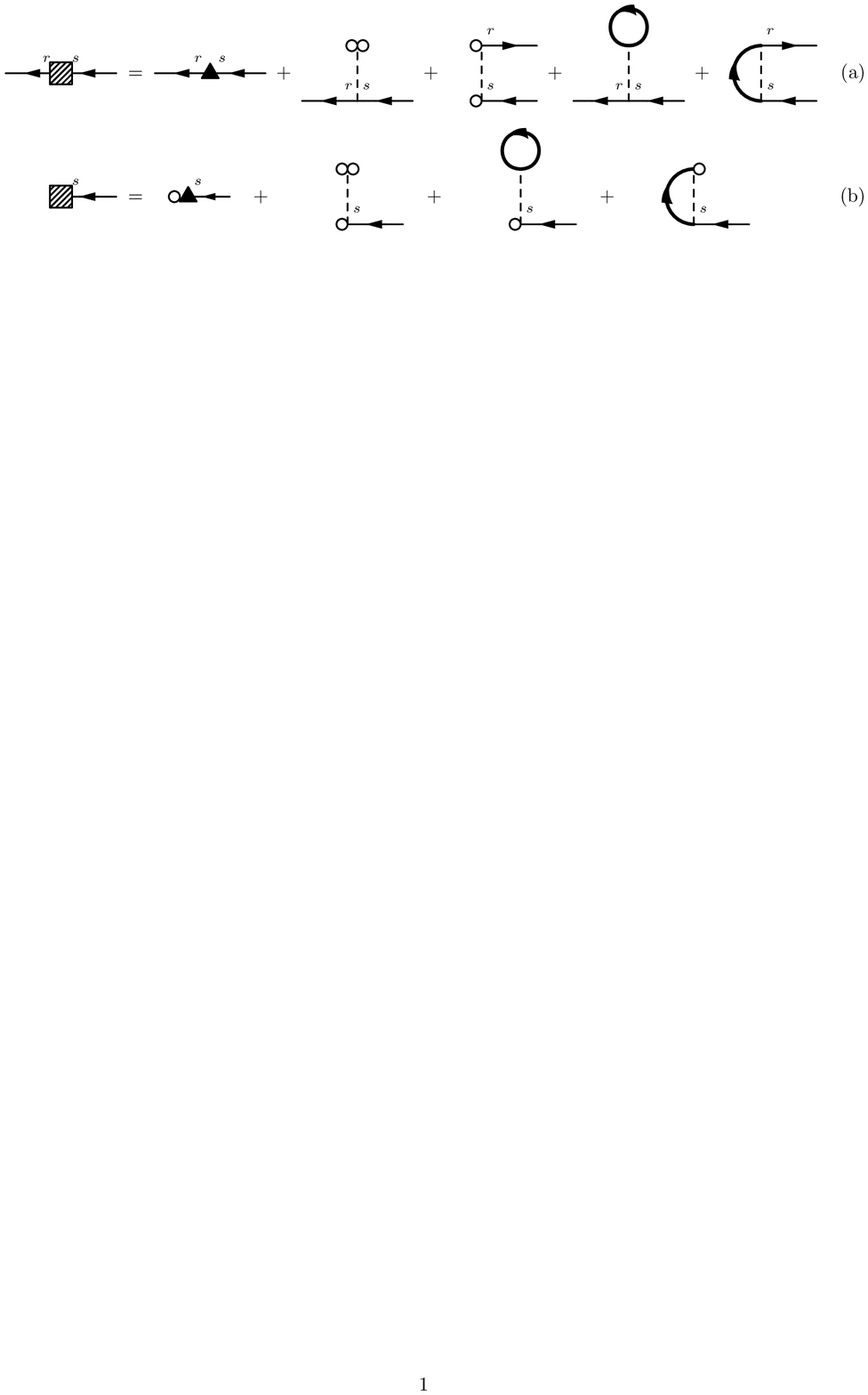}
  \caption{The Feynman diagrams contributing to the self-energy
    $\widehat\Sigma^{rs}_{11}$ of the internal line (a) and the
    contribution of the tadpole diagrams (b). The heavy line depicts
    the internal line, while the thin line corresponds to the free
    propagator.  The circle denotes a condensate operator, the dashed
    line is for the interaction and the triangle corresponds to
    $-\mu$.}
  \label{fig:intse}
\end{figure}
We denote the Hartree-Fock propagators (which will play the role of
internal lines) by $\widehat\Gr^{rs}_{11}\komega$ and define it with
the following equation:
\begin{equation}
  \label{eq:intline}
  \widehat\Gr^{rs}_{11}\komega=\frac{\delta_{rs}}{i\omega_n-\hslash^{-1}\kink-
    \widehat\Sigma^{rs}_{11}},
\end{equation}
where $\widehat\Sigma^{rs}_{11}$ is the momentum and frequency
independent self-energy of the internal line. The Feynman diagrams
contributing to the self-energy of the internal line can be seen in
Fig.  \ref{fig:intse}. The corresponding equation reads as:
\begin{equation}
  \label{eq:intse}
  \widehat\Sigma^{rs}_{11}=\hslash^{-1}\left[-\mu\delta_{rs}+\left(N_0 \zeta_{r'}
  \zeta_{s'}+H^{r's'}\right)\left(V^{r's'}_{rs}+V^{rr'}_{s's}\right)\right],
\end{equation}
where $H^{r's'}$ is the contribution of the loop (appearing in the
Hartree and Fock terms):
\begin{equation}
  \label{eq:hartree}
  H^{rs}=\lim_{\eta\to0}\int\frac{d^3q}{(2\pi)^3}
  \frac{1}{\beta\hslash}\sum_{i\nu_n}\left[-\widehat\Gr^{rs}_{11}(\vec{q},i\nu_n)
  \right]e^{i\nu_n\eta}
  =\frac{\delta_{rs}}{(2\pi)^2\lambda^3}\Gamma(3/2)
  F(3/2,\beta\hslash\widehat\Sigma^{rr}_{11}),
\end{equation}
with
\begin{equation}
  \label{eq:twl}
  \lambda=\frac{\hslash}{\sqrt{2Mk_BT}}
\end{equation}
the thermal wavelength, $\Gamma(s)$ the gamma function and
$F(s,\gamma)$ the Bose-Eintein integral with parameter $s$ and
argument $\gamma$ \cite{Robinson}.

The value of the chemical potential is determined by the requirement
Eq. \eqref{eq:vanelexc}, or equivalently by the vanishing of the
contribution of tadpole diagrams \cite{SzSz2}:
\begin{equation}
  \label{eq:tadpole}
  0=\Sigma^s_{01}=\hslash^{-1}\sqrt{N_0}\zeta_r\Big[-\mu\delta_{rs}
    + N_0\zeta_{r'}\zeta_{s'}V^{r's'}_{rs}
    + H^{r's'}\left(V^{r's'}_{rs}+V^{r'r}_{s's}\right)\Big].
\end{equation}

Since we work in the ordered phase, $N_0\neq0$ everywhere assumed.
The simultaneous solution of equations \eqref{eq:intline},
\eqref{eq:intse}, \eqref{eq:hartree}, \eqref{eq:tadpole} (which is the
consequence of Eq. \eqref{eq:vanelexc}), and Eq. \eqref{eq:eqstate1}
provides the self-energy and propagator of the internal lines.

For a \textit{ferromagnetic system} the condensate spinor can be
chosen to be $\zeta_r=(1,0,0)^T$. According to Eq. \eqref{eq:tadpole}
the chemical potential for this case reads as:
\begin{equation}
  \label{eq:fchp}
  \mu=c_n \big(N_0+2H^{++}+H^{00}+H^{--}\big)
  +c_s\big(N_0+2H^{++}+H^{00}-H^{--}\big),
\end{equation}
while from Eqs. \eqref{eq:intse} and \eqref{eq:fchp} the self-energies
take the following form:
\begin{subequations}
  \label{eqs:seif}
  \begin{align}
    \hslash\widehat\Sigma^{++}_{11}&=(c_n+c_s)N_0,\\
    \hslash\widehat\Sigma^{00}_{11}&=-(c_n+c_s)H^{++}+(c_n-c_s)H^{00}+2c_sH^{--},\\
    \hslash\widehat\Sigma^{--}_{11}&=-(c_n+3c_s)H^{++}+(c_n+3c_s)H^{--}-2c_sN_0.
  \end{align}
\end{subequations}
The contributions of the single loops are also related to the
self-energies through Eq.  \eqref{eq:hartree}. There is a final
equation from Eq. \eqref{eq:eqstate1} reading as:
\begin{equation}
  \label{eq:es1HF}
  N=N_0+H^{++}+H^{00}+H^{--}.
\end{equation}
The system of equations \eqref{eq:fchp}, \eqref{eqs:seif},
\eqref{eq:hartree} and \eqref{eq:es1HF} form the equation of state in
the following way. One can express the condensate density with the
help of $N,H^{++},H^{00},H^{--}$ from Eq.  \eqref{eq:es1HF} and
substitute it to Eqs. \eqref{eqs:seif}. These equations now form a
closed set of equations toether with Eq. \eqref{eq:hartree}. One can
solve them for $H^{++},H^{00},H^{--}$ as functions of $T,N$, and
substitute them to Eq. \eqref{eq:fchp} to arrive at the equation of
state $\mu=\mu(T,N)$. An alternate way is to express $N_0$ from Eq.
\eqref{eq:fchp} and solve the equations for the loop terms with
$T,\mu$ fixed, and then express $N=N(T,\mu)$. The two equations of
state are necessarily equivalent.

For a \textit{polar system} the condensate spinor can be taken as
$\zeta_r=(0,1,0)^T$. The chemical potential for this phase, from Eq.
\eqref{eq:tadpole}, reads as:
\begin{equation}
  \label{eq:pchp}
  \mu=c_n \big(N_0+H^{++}+2H^{00}+H^{--}\big)
  +c_s\big(H^{++}+H^{--}\big).
\end{equation}
The self-energies \eqref{eq:intse} in this case take the following
form:
\begin{subequations}
  \label{eqs:seip}
  \begin{align}
    \hslash\widehat\Sigma^{00}_{11}&=c_nN_0,\\
    \hslash\widehat\Sigma^{++}_{11}&=(c_n-c_s)(H^{++}-H^{00})+c_s N_0,
  \end{align}
\end{subequations}
where we have used the specific form of the chemical potential
\eqref{eq:pchp} and that
$\widehat\Sigma^{++}_{11}=\widehat\Sigma^{--}_{11}$ and
$H^{++}=H^{--}$. Equation \eqref{eq:eqstate1} can be expressed as:
\begin{equation}
  \label{eq:es1HFp}
  N=N_0+2 H^{++}+H^{00}.
\end{equation}
Expressing $N_0$ from Eq. \eqref{eq:es1HFp} and substituting it to
Eqs. \eqref{eqs:seip} the latter ones can be solved together with Eq.
\eqref{eq:hartree} for $H^{00}$ and $H^{++}$. They can be used in Eq.
\eqref{eq:pchp} to express the equation of state as $\mu=\mu(T,N)$.
The alternate way is also possible here, i.e. to express the total
number of particles as $N=N(T,\mu)$.

\subsection{Effective interactions}
\label{ssec:effint}

In this subsection we discuss the effective interactions which will
play an important role in the following. The self consistent solution
of the Hartree-Fock propagators contains a partial summation of
infinite number of Feynman graphs. This partial summation is visible
from the iterative solution of the Hartree-Fock propagators. With
iteration one arrives to effective interactions which are the
contributions of repeated exchange and direct terms.

\subsubsection{Repeated exchange interactions}

\begin{figure}[!ht]
  \centering
  \includegraphics*[67mm,260mm][151mm,278mm]{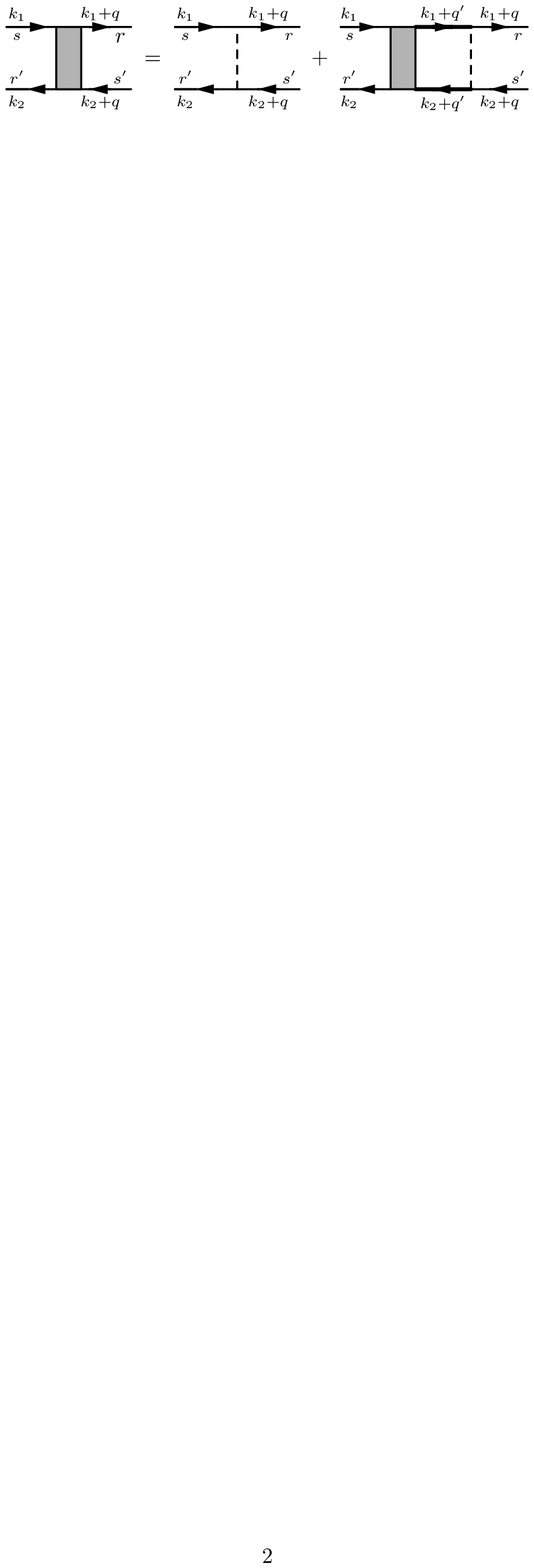}
  \caption{The Feynman diagrams defining $\Theta^{rs}_{r's'}(k_1,k_2,q)$.
    The shaded square depicts $\Theta$.}
  \label{fig:fdefrt}
\end{figure}

Denote the contribution of multiple exchange interactions with the
quantity $\Theta^{rs}_{r's'}(k_1,k_2,q)$, defined by the
Feynman-graphs in Fig. \ref{fig:fdefrt}.
For the sake of brevity one can introduce the four-vector notation:
$q=(\vec{q},i\nu_n)$ and $k=(\vec{k},i \omega_n)$. The equation for
$\Theta$ reads as:
\begin{equation}
  \label{eq:thetaeq}
  \Theta^{rs}_{r's'}(k_1,k_2,q) = V^{rs}_{r's'} + \int Dq'\Big[
  \Theta^{as}_{r'b}(k_1,k_2,q')\, {\rho_{(0)}}^{ba}_{cd}(k_1+q',k_2+q')\,
  V^{rc}_{ds'}\Big],
\end{equation}
where
\begin{subequations}
  \begin{align}
    \int Dk&\equiv\frac{1}{\beta\hslash}\sum_{i\omega_n}\int
    \frac{d^3 k}{(2\pi)^3}, \quad \text{and }\\
    {\rho_{(0)}}^{s_2 s_1}_{r_1 r_2}(k_1, k_2) & := -\hslash^{-1}
    \widehat\Gr^{s_1 r_1}_{11}(k_1) \widehat\Gr^{s_2 r_2}_{11}(k_2)
  \end{align}
\end{subequations}
were also introduced. The contribution of the bubble graph can be
calculated from $\rho_{(0)}$, which we give here for future reference:
\begin{equation}
  \label{eq:bubfromr}
  \int Dq \; {\rho_{(0)}}^{s_2 s_1}_{r_1 r_2}(q, k+q) =
  {\Pi_{(0)}}^{s_2 s_1}_{r_1 r_2}(k).
\end{equation}

Due to the conservation of the spin \textrm{z} component Eq.
\eqref{eq:thetaeq} decouples to independent matrix equations. Let us
define the following matrices for 0 spin transfer:
\begin{subequations}
  \label{eqs:theta0}
  \begin{align}
    \mat[0][\Theta]&=\left[\begin{array}{c c c}
        \Theta^{++}_{++} & \Theta^{0+}_{+0} & \Theta^{-+}_{+-}\\
        \Theta^{+0}_{0+} & \Theta^{00}_{00} & \Theta^{-0}_{0-}\\
        \Theta^{+-}_{-+} & \Theta^{0-}_{-0} & \Theta^{--}_{--}
      \end{array}\right],\\
    \mat[0][V]&=
    \left[\begin{array}{c c c}
        c_n+c_s & c_s & 0\\
        c_s & c_n & c_s \\
        0 & c_s & c_n+c_s
      \end{array}\right],\\
    \mat[0][\rho]\nul&= \left[ \begin{array}{c c c}
        {\rho_{(0)}}^{++}_{++} & 0 & 0\\
        0 & {\rho_{(0)}}^{00}_{00} & 0\\
        0 & 0 & {\rho_{(0)}}^{--}_{--}
        \end{array}\right].
  \end{align}
\end{subequations}
For +1 spin transfer one defines
\begin{subequations}
    \label{eqs:theta+}
  \begin{align}
    \mat[+][\Theta]&=\left[\begin{array}{c c}
        \Theta^{++}_{00} & \Theta^{0+}_{0-}\\
        \Theta^{+0}_{-0} & \Theta^{00}_{--}
        \end{array}\right],\\
    \mat[+][V]&=\left[\begin{array}{c c}
        c_n & c_s\\
        c_s & c_n
        \end{array}\right],\\
      \mat[+][\rho]\nul&=\left[\begin{array}{c c}
        {\rho_{(0)}}^{0+}_{+0} & 0\\
        0 & {\rho_{(0)}}^{-0}_{0-}
        \end{array}\right].
  \end{align}
\end{subequations}
And finally, for +2 spin transfer one introduces
\begin{subequations}
  \label{eqs:thetaQ}
  \begin{align}
    \mat[Q][\Theta]&=\Theta^{++}_{--},\\
    \mat[Q][V]&=c_n-c_s,\\
    \mat[Q][\rho]\nul&={\rho_{(0)}}^{-+}_{+-}.
  \end{align}
\end{subequations}
With the definitions \eqref{eqs:theta0}, \eqref{eqs:theta+} and
\eqref{eqs:thetaQ} one can cast Eq. \eqref{eq:thetaeq} to the
following integral equations for $n\in\{0,+,Q\}$:
\begin{equation}
  \label{eq:inteqm}
  \int Dq' \; \mat[n][\Theta](k_1,k_2,q')\Big[\delta(q-q') \mat[][1]
  -\mat[n][\rho]\nul(k_1+q',k_2+q') \mat[n][V]\Big]=\mat[n][V].
\end{equation}
The $\delta$-function is normalized to unity with the definition
\begin{equation}
  \label{eq:deltanorm}
  \delta(k)\equiv(2\pi)^3 (\beta\hslash) \delta^{(3)}(\vec{k})
  \delta_{\omega_n,0}.
\end{equation}
It is straightforward to check that the solution of Eq.
\eqref{eq:inteqm} reads as:
\begin{subequations}
  \label{eqs:thetasols}
  \begin{equation}
    \label{eq:thetasol}
    \mat[n][\Theta](k_1,k_2,q)=\mat[n][V] \c\epss{n}^{-1}(k_2-k_1),
  \end{equation}
  with
  \begin{equation}
    \label{eq:dielstar}
    \epss{n}(k)=\mat[][1]-\mat[n][\Pi]\nul(k) \mat[n][V].
  \end{equation}
\end{subequations}
The matrix $\mat[n][\Pi]\nul$ is understood as the integral of the
matrix $\mat[n][\rho]\nul$ according to Eq. \eqref{eq:bubfromr}. The
matrices $\epss{n}$ can be interpreted as dielectric functions with
respect to repeated exchange interactions. It is important to note
that $\mat[n][\Theta](k_1,k_2,q)$ does not depend on the amount of
momentum transferred ($q$) at all, as is visible from Eq.
\eqref{eq:thetasol}.

\subsubsection{Repeated direct interactions}

\begin{figure}[!ht]
  \centering
  \includegraphics*[51mm,260mm][165mm,278mm]{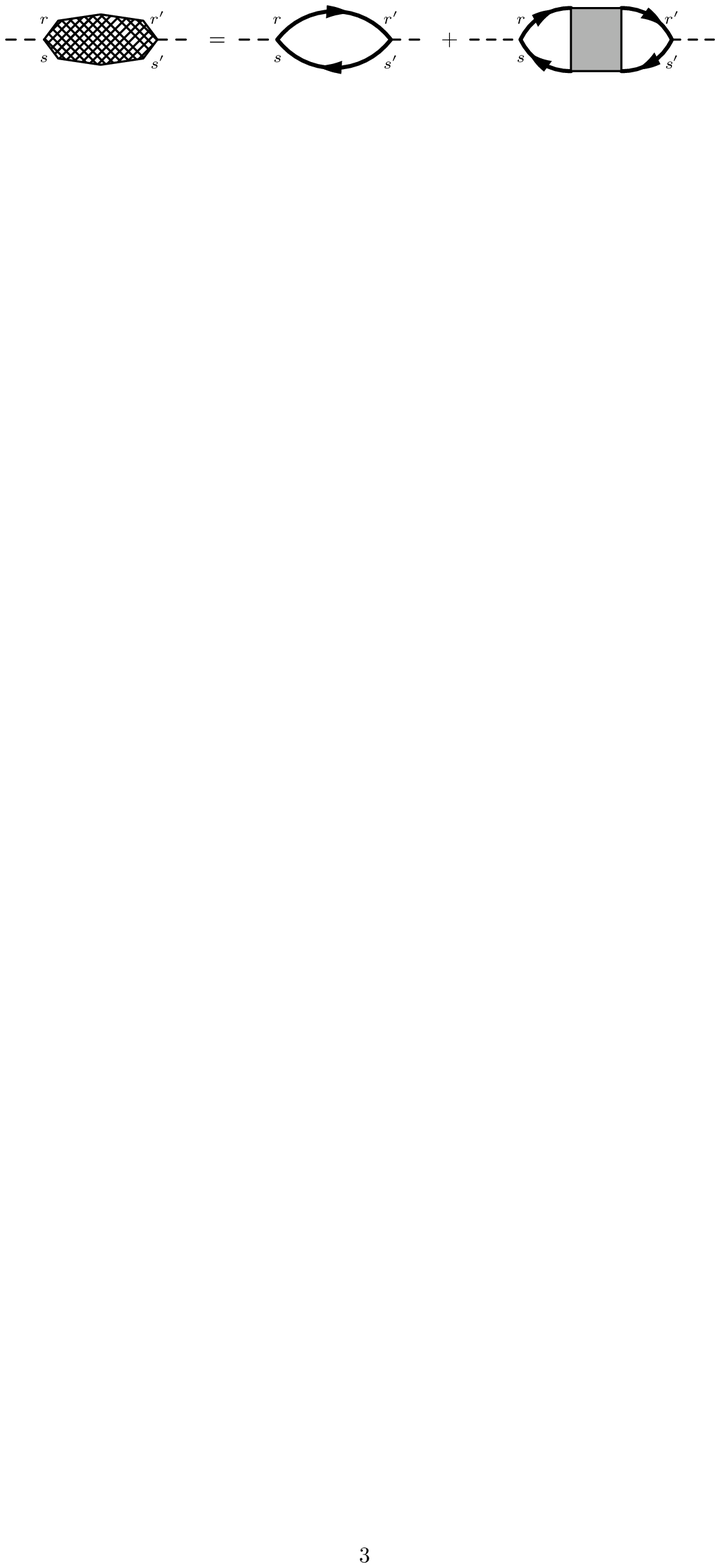}
  \caption{The Feynman diagrams defining the regular polarization
    $\Pi^{(r)sr}_{r's'}$. The hatched poligon represents the regular
    polarization function.}
  \label{fig:regpol}
\end{figure}

For the determination of the effective interaction representing
multiple direct interactions, first consider the regular polarization
functions, defined by the Feynman diagrams depicted in Fig.
\ref{fig:regpol}, or with the following equation (already separated by
spin transfer):
\begin{equation}
  \label{eq:regpole}
  \mat[n][\Pi]^{(r)}(k)=\mat[n][\Pi]\nul(k)+\int Dq\; Dq'\;\mat[n][\rho]\nul
  (q,k+q)\mat[n][\Theta](q,k+q,q'-q)\;\mat[n][\rho]\nul(q',k+q'),
\end{equation}
which leads to
\begin{equation}
  \label{eq:regpols}
  \mat[n][\Pi]^{(r)}(k)=\epss{n}^{-1}(k)\mat[n][\Pi]\nul(k).
\end{equation}

\begin{figure}[!ht]
  \centering
  \includegraphics*[70mm,240mm][148mm,278mm]{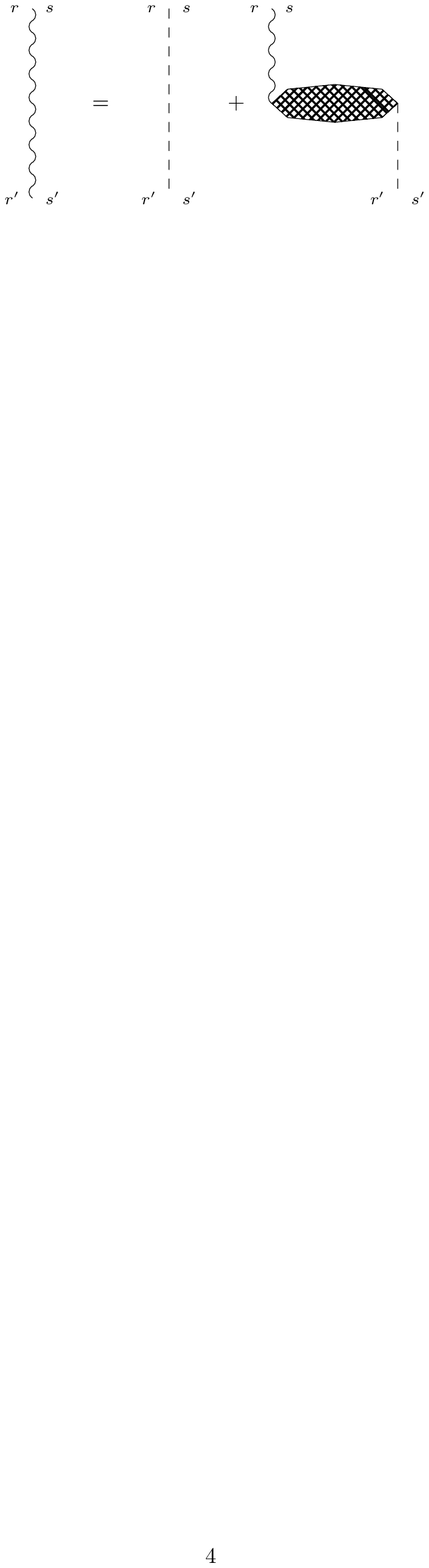}
  \caption{The Feynman diagrams defining the direct effective interaction. The
    wiggly line represents the direct effective interaction.}
  \label{fig:direffint}
\end{figure}

The direct effective interaction is then defined by the Feynman
diagrams in Fig. \ref{fig:direffint}, or by the equation
\begin{equation}
  \label{eq:direffint}
  W^{rs}_{r's'}(k)=V^{rs}_{r's'} + W^{rs}_{ab}(k) \; \Pi^{(r)ba}_{cd}(k)\;
  V^{dc}_{r's'}.
\end{equation}
Equation \eqref{eq:direffint} splits into separate matrix equations
according to the amount of spin transferred. Introducing the
quantities:
\begin{subequations}
  \label{eqs:strwc0}
  \begin{align}
    \mat[0][W]&=\left[\begin{array}{c c c}
        W^{++}_{++} & W^{++}_{00} & W^{++}_{--}\\
        W^{00}_{++} & W^{00}_{00} & W^{00}_{--}\\
        W^{--}_{++} & W^{--}_{00} & W^{--}_{--}
      \end{array}\right],
    &\mat[0][C]&=\left[\begin{array}{c c c}
        c_n+c_s & c_n & c_n-cs\\
        c_n & c_n & c_n \\
        c_n-c_s & c_n & c_n+c_s
        \end{array}\right],\\
      \mat[+][W]&=\left[\begin{array}{c c}
          W^{0+}_{+0} & W^{0+}_{0-}\\
          W^{-0}_{+0} & W^{-0}_{0-}
        \end{array}\right],
      &\mat[+][C]&=\left[\begin{array}{c c}
          c_s & c_s\\
          c_s & c_s
        \end{array}\right],\\
      \mat[Q][W]&=W^{-+}_{+-}, &\mat[Q][C]&=0,
  \end{align}
\end{subequations}
the resulting equations are
\begin{equation}
  \label{eq:direffint2}
  \mat[n][W](k)=\mat[n][C]+\mat[n][W](k) \; \mat[n][\Pi]^{(r)}(k)\;
  \mat[n][C]
\end{equation}
for $n\in\{0,+,Q\}$. From $\mat[Q][C]=0$ it immediately follows that
$\mat[Q][W]=0$.

The solutions of Eq. \eqref{eq:direffint2} reads as:
\begin{equation}
  \label{eq:deisol1}
  \mat[n][W](k)=\mat[n][C]\; {\mat[n][\eps]^{(r)}}^{-1}(k),
\end{equation}
with
\begin{equation}
  \label{eq:deisol2}
  \mat[n][\eps]^{(r)}=\mat[n][1]-\mat[n][\Pi]^{(r)}\;\mat[n][C].
\end{equation}

\subsubsection{The equation of state in terms of the effective interactions}

\begin{figure}[!ht]
  \centering
  \includegraphics*[30mm,146mm][190mm,278mm]{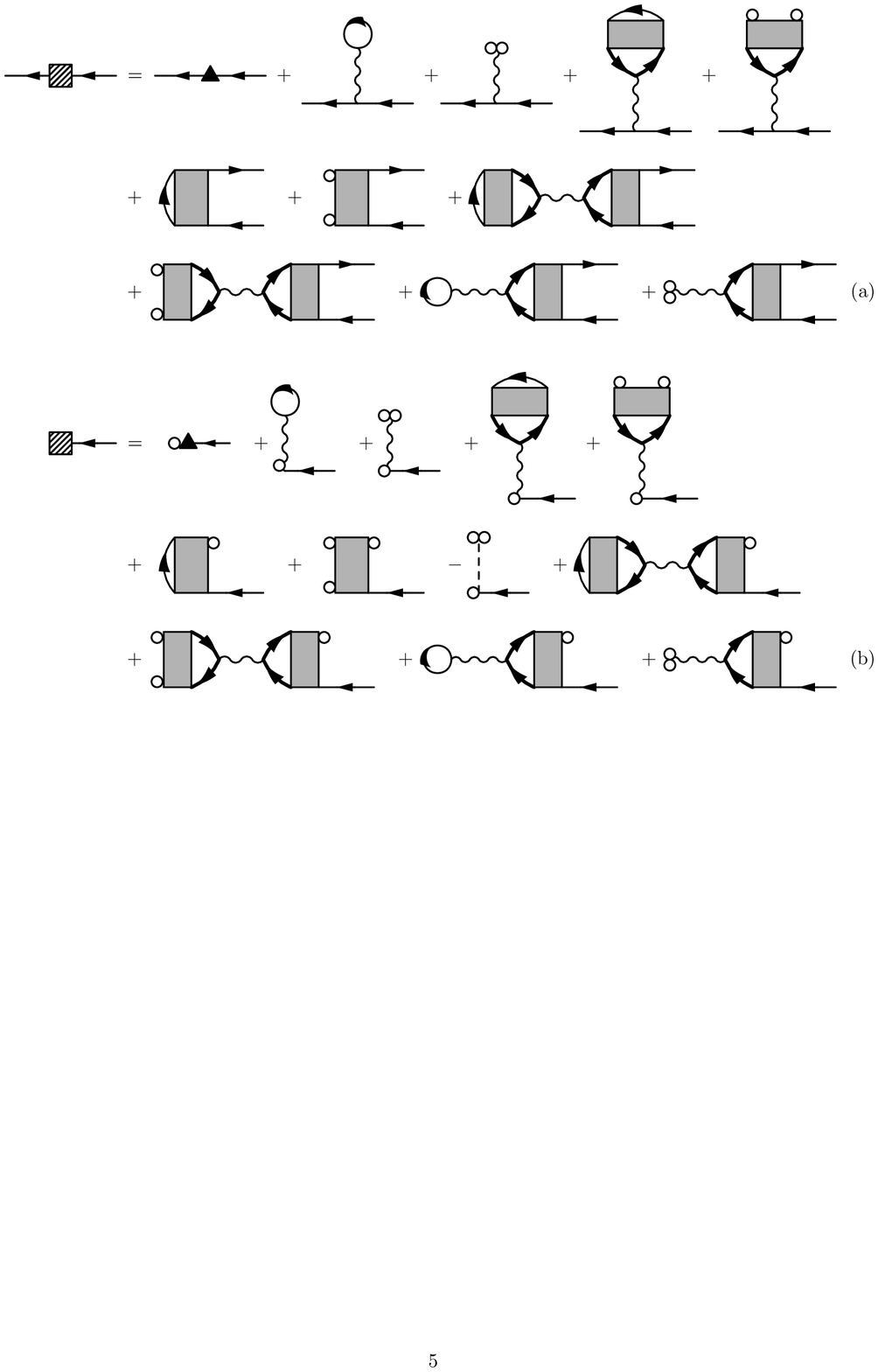}
  \caption{The Feynman diagrams contributing to the self-energy 
    $\widehat\Sigma^{rs}_{11}$ of the internal line (a), and to the
    tadpole diagrams (b).}
  \label{fig:HFrev}
\end{figure}

The final structure of the Hartree-Fock self-energies and tadpole
diagrams, as a result of the iteration of the Hartree-Fock propagator,
are depicted in Fig. \ref{fig:HFrev}. a) and b), respectively, with
the help of the effective interactions.
For the sake of simplicity we have omitted the spin indices from the
figure. Note, that in the final structure of the self-energies in Fig.
\ref{fig:HFrev}. a) the Hartree-Fock propagators do not join vertices
of the same interaction line, such propagators are always unperturbed
ones. The iteration, of course, can be continued by substituting the
remaining Hartree-Fock propagators with self-energies given in Fig.
\ref{fig:intse} a) in the self-energies, a procedure without end.
These further iterations, however, do not generate new blocks, i.e.
the building blocks remain the effective interactions $\Theta$ and
$W$.

\subsubsection{Anomalous vertex functions}
\label{sssec:avf}

For the sake of a simplified notation it is convenient to introduce
the anomalous vertex functions $\widetilde\Lambda^{sr}_{a\alpha}(k)$,
and $\widetilde\Lambda^{a\alpha}_{r's'}(k)$ , whose general
definitions can be found in \cite{SzSz2}.  In the Hartree-Fock
approximation the Feynman diagrams for $\widetilde\Lambda^{sr}_{a,1}$
are depicted in Fig. \ref{fig:pav}.  The equations for the anomalous
vertices in this approximation read as:
\begin{figure}[!hb]
  \centering
  \includegraphics*[62mm,260mm][148mm,278mm]{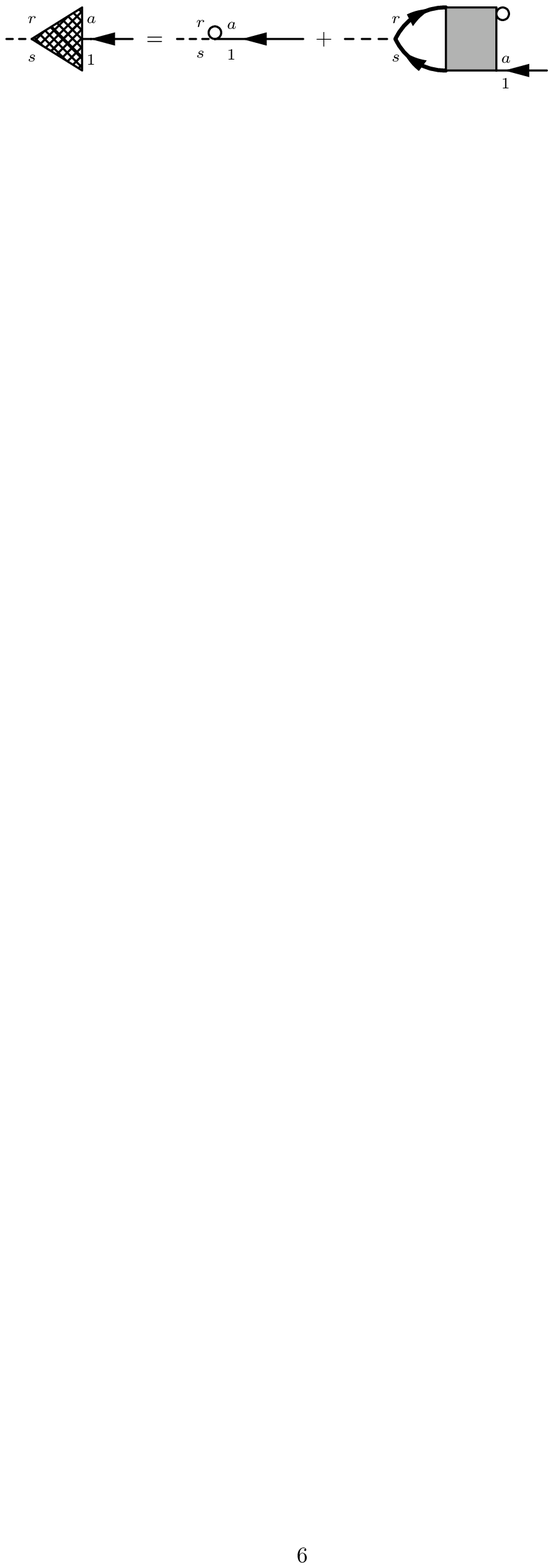}
  \caption{The Feynman diagrams defining the proper anomalous vertex. The
    hatched triangle represents the proper anomalous vertex.}
  \label{fig:pav}
\end{figure}
\begin{subequations}
  \label{eqs:pavge}
  \begin{align}
    \widetilde\Lambda^{sr}_{a,1}(k)&=\Lambda^{sr}_{(0)a,1} + \int Dq\;
    \rho^{sr}_{(0)cd}(q,k+q)\Theta^{r'c}_{da}(q,k+q,-q)
    \Lambda^{s'r'}_{(0)a,1},\quad \text{with}\quad \Lambda^{sr}_{(0)a,1}
    =\delta_{s,a}\zeta_r\sqrt{N_0},\label{eq:pavge1}\\
    \widetilde\Lambda^{sr}_{a,-1}(k)&=\Lambda^{sr}_{(0)a,-1} + \int Dq\;
    \rho^{sr}_{(0)cd}(q,k+q)\Theta^{r'c}_{da}(q,k+q,-k-q)
    \Lambda^{s'r'}_{(0)a,-1},\quad \text{with}\quad \Lambda^{sr}_{(0)a,-1}
    =\delta_{r,a}\zeta_s\sqrt{N_0},\label{eq:pavge2}\\
    \widetilde\Lambda^{a,1}_{r's'}(k)&=\Lambda^{a,1}_{(0)r's'} + \int Dq\;
    \Lambda^{a,1}_{(0)rs}\Theta^{cr}_{sd}(0,k,q)\rho^{dc}_{(0)r's'}(q,k+q),
    \quad \text{with}\quad\Lambda^{a,1}_{(0)r's'}=\delta_{a,s'}
    \zeta_{r'}\sqrt{N_0},\label{eq:pavge3}\\
    \widetilde\Lambda^{a,-1}_{r's'}(k)&=\Lambda^{a,-1}_{(0)r's'} + \int Dq\;
    \Lambda^{a,-1}_{(0)rs}\Theta^{cr}_{sd}(-k,0,k+q)\rho^{dc}_{(0)r's'}(q,k+q),
    \quad \text{with}\quad\Lambda^{a,-1}_{(0)r's'}=\delta_{a,r'}
    \zeta_{s'}\sqrt{N_0}.\label{eq:pavge4}
  \end{align}
\end{subequations}
Note that in the diagrams for \eqref{eq:pavge3} and \eqref{eq:pavge4}
the type of left and right connections are interchanged as compare to
the diagrams for \eqref{eq:pavge1} and \eqref{eq:pavge2},
respectively.

Equations \eqref{eqs:pavge} also split to separate equations according
to the amount of spin transfered from the right to the left.  Let us
define
\begin{subequations}
  \label{eqs:avvp}
  \begin{align}
    \vv[][0][\widetilde\Lambda]_{\alpha}&=\left(\begin{array}{c}
        \widetilde\Lambda^{++}_{0\alpha}\\
        \widetilde\Lambda^{00}_{0\alpha}\\
        \widetilde\Lambda^{--}_{0\alpha}
      \end{array}\right), &
    \vv[][0][\widetilde\Lambda]^{\alpha}&=\left(\begin{array}{c}
        \widetilde\Lambda^{0\alpha}_{++}\\
        \widetilde\Lambda^{0\alpha}_{00}\\
        \widetilde\Lambda^{0\alpha}_{--}
      \end{array}\right),\label{eq:vectvert0pol}\\
    \vv[][+][\widetilde\Lambda]_{1}&=\left(\begin{array}{c}
        \widetilde\Lambda^{0+}_{-,1}\\
        \widetilde\Lambda^{-0}_{-,1}
      \end{array}\right), &
    \vv[][+][\widetilde\Lambda]^{1}&=\left(\begin{array}{c}
        \widetilde\Lambda^{-,1}_{+0}\\
        \widetilde\Lambda^{-,1}_{0-}
      \end{array}\right), \nonumber \\
    \vv[][+][\widetilde\Lambda]_{-1}&=\left(\begin{array}{c}
        \widetilde\Lambda^{0+}_{+,-1}\\
        \widetilde\Lambda^{-0}_{+,-1}
      \end{array}\right), &
    \vv[][+][\widetilde\Lambda]^{-1}&=\left(\begin{array}{c}
        \widetilde\Lambda^{+,-1}_{+0}\\
        \widetilde\Lambda^{+,-1}_{0-}
      \end{array}\right), \label{eq:vectvert+pol}
  \end{align}
\end{subequations}
for the polar case and
\begin{subequations}
  \label{eqs:avvf}
  \begin{align}
    \vv[][0][\widetilde\Lambda]_{\alpha}&=\left(\begin{array}{c}
        \widetilde\Lambda^{++}_{+\alpha}\\
        \widetilde\Lambda^{00}_{+\alpha}\\
        \widetilde\Lambda^{--}_{+\alpha}
      \end{array}\right), &
    \vv[][0][\widetilde\Lambda]^{\alpha}&=\left(\begin{array}{c}
        \widetilde\Lambda^{+\alpha}_{++}\\
        \widetilde\Lambda^{+\alpha}_{00}\\
        \widetilde\Lambda^{+\alpha}_{--}
      \end{array}\right),\label{eq:vectvert0fer}\\
    \vv[][+][\widetilde\Lambda]_{1}&=\left(\begin{array}{c}
        \widetilde\Lambda^{0+}_{0,1}\\
        \widetilde\Lambda^{-0}_{0,1}
      \end{array}\right), &
    \vv[][+][\widetilde\Lambda]^{1}&=\left(\begin{array}{c}
        \widetilde\Lambda^{0,1}_{+0}\\
        \widetilde\Lambda^{0,1}_{0-}
      \end{array}\right), \nonumber \\
    \vv[][+][\widetilde\Lambda]_{-1}&=\vv[][][0], &
    \vv[][+][\widetilde\Lambda]^{-1}&=\vv[][][0], \label{eq:vectvert+fer}\\
    \vv[][Q][\widetilde\Lambda]_{1}&=\widetilde\Lambda^{-+}_{-,1}, &
    \vv[][Q][\widetilde\Lambda]^{1}&=\widetilde\Lambda^{-,1}_{+-}, \nonumber\\
    \vv[][Q][\widetilde\Lambda]_{-1}&=0, &
    \vv[][Q][\widetilde\Lambda]^{-1}&=0,
  \end{align}
\end{subequations}
for the ferromagnetic one. The resulting equation for
$\vv[][n][\widetilde\Lambda]_\alpha$ reads as:
\begin{equation}
  \label{eq:pave}
  \vv[][n][\widetilde\Lambda]_{\alpha}(k)=\vv[][n][\Lambda]_{(0)\alpha}
  +\int Dq\,\mat[n][\rho]\nul(q,k+q)\,\mat[n][\Theta](q,k+q,-q)\,
  \vv[][n][\Lambda]_{(0)\alpha}
\end{equation}
for $\alpha=\pm1$. Where the $\vv[][][\Lambda]_{(0)}$ vectors are
defined similarly as the $\vv[][][\widetilde\Lambda]$ vectors in
Eqs. \eqref{eqs:avvp} and \eqref{eqs:avvf}. We have used the
consequence of Eq. \eqref{eq:thetasol}, namely that $\Theta$ does not
depend on its third argument. The solution of Eq. \eqref{eq:pave}
reads as:
\begin{equation}
  \label{eq:pav1}
  \vv[][n][\widetilde\Lambda]_\alpha(k)=\epss{n}^{-1}(k)\,
  \vv[][n][\Lambda]_{(0)\alpha}.
\end{equation}
Similar considerations can be carried on to
$\vv[][n][\widetilde\Lambda]^\alpha(k)$ with the result:
\begin{equation}
  \label{eq:pav2}
  \vv[][n][\widetilde\Lambda]^\alpha(k)=\epss{n}^{-1}(k)\,
  \vv[][n][\Lambda]_{(0)}^\alpha.
\end{equation}

\subsection{Self-energies in the Hartree-Fock approximation}
\label{ssec:seHF}

Until now we have considered the Hartree-Fock propagators, which are
used in the equation of state. The thermal Green's function of the
system, defined by Eq. \eqref{eq:grdef} and expressed with the help of
the self-energies by the generalized Dyson-Beliaev equation
\eqref{eq:dyson}, are more complicated. To be consistent both with the
partial summation made in the equation of state and the consequences
of symmetry breaking, one has to construct the self-energies from the
tadpole diagrams in Fig. \ref{fig:HFrev}. b) by changing a circle
(representing a condensate atom) to an appropriate particle line. The
self-energies obtained are diagonal in the upper (Roman) indices;
moreover this index is $0$ in the polar and $+$ in the ferromagnetic
case, i.e. agrees with that of the condensate. The self-energies
$\Sigma^{00}_{11}$ and $\Sigma^{++}_{11}$ satisfy by construction the
relationship:
\begin{equation}
  \label{eq:HP}
  \frac{1}{\sqrt{N_0}}\Sigma^c_{0,1}=\Sigma^{cc}_{11}(0,0)-
  \Sigma^{cc}_{-1,1}(0,0),
\end{equation}
where $c=0$ for the polar and $c=+$ for the ferromagnetic case. The
Eq. \eqref{eq:HP} amounts to a generalization of the Hugenholtz-Pines
theorem \cite{HP} to the spin-1 systems, since
\begin{equation}
  \label{eq:HPexp}
  \Sigma^c_{0,1}=0, \quad \text{so} \quad \Sigma^{'cc}_{11}(0,0)-
  \Sigma^{cc}_{-1,1}(0,0)=\mu,
\end{equation}
where $\Sigma'$ is the self-energy without the subtraction $\mu$. This
ensures that the corresponding excitations are gapless as required by
the general Goldstone theorem due to the breaking of the gauge
symmetry.

As for the other self-energies consistency requires that they should
be built by using the same diagrammatic blocks and to the same order
in them. The resulting Feynman diagrams are depicted in Fig.
\ref{fig:nse1}. a) and b). The more detailed diagrammatic equation is
shown in appendix \ref{sec:sed}. (Note that in the ferromagnetic phase
$\Sigma^{rs}_{-1,1}(k)=0$ except when $r=s=+$ due to the structure of
the interaction). One expects that $\Sigma^{00}_{11}(0,0)=0$ in the
ferromagnetic phase and
$\Sigma^{--}_{11}(0,0)-\Sigma^{+-}_{-1,1}(0,0)=0$ in the polar phase.
These relationships lead to gapless spin wave spectra as required by
the Goldstone theorem, since the rotational symmetry in the spin space
is also broken, besides the gauge symmetry. The spectra to be
calculated in the next section will exhibit this property.

According to Fig. \ref{fig:nse1}. a) the self-energy contributions can
be brought into two categories. Those graphs are in the first
category, which can not be split into two (connected to external
lines) by cutting a single interaction line. The first four diagrams
fall under this category. We call these self-energy contributions
proper.  The second category contains those graphs, which do not
exhibit this property. The last diagram in Fig.  \ref{fig:nse1}. a)
falls under this category. We call these latter self-energy
contributions improper.  Therefore the equation of the normal
self-energies read as:
\begin{subequations}
  \label{eqs:se1}
  \begin{align}
    \Sigma^{rs}_{11}(k)&=\widetilde\Sigma^{rs}_{11}(k)+M^{rs}_{11}(k),
    \quad \text{with}\\
    \widetilde\Sigma^{rs}_{11}(k) &= \hslash^{-1}\bigg[-\mu + N_0
    \Theta^{s'r'}_{rs}(0,k,0)+H^{r's'}\Big(V^{r's'}_{rs}+V^{r'r}_{s's}\Big)
    \bigg],\label{eq:pse1}\\
    M^{rs}_{11}(k)&=\hslash^{-1}\widetilde\Lambda^{r,1}_{cd}(k)W^{dc}_{ef}(k)
    \widetilde\Lambda^{fe}_{s,1}(k).\label{eq:ipse1}
  \end{align}
\end{subequations}
where tilde refers to proper diagrams as in Subsection \ref{sssec:avf}.
The resulting diagrams for $\Sigma^{rs}_{-1,1}$ are seen in Fig.
\ref{fig:nse1}. b) and read as:
\begin{subequations}
  \label{eqs:se2}
  \begin{align}
    \Sigma^{rs}_{-1,1}(k)&=\widetilde\Sigma^{rs}_{-1,1}(k)+M^{rs}_{-1,1}(k),
    \quad \text{with}\\
    \widetilde\Sigma^{rs}_{-1,1}(k) &= \hslash^{-1} N_0\Big[
    \Theta^{s'r}_{r's}(-k,0,k)-V^{s'r}_{r's}\Big],\label{eq:pse2}\\
    M^{rs}_{-1,1}(k)&=\hslash^{-1}\widetilde\Lambda^{r,-1}_{cd}(k)
    W^{dc}_{ef}(k)\widetilde\Lambda^{fe}_{s,1}(k).\label{eq:ipse2}
  \end{align}
\end{subequations}

\begin{figure}[t!]
  \centering
  \includegraphics*[30mm,216mm][190mm,278mm]{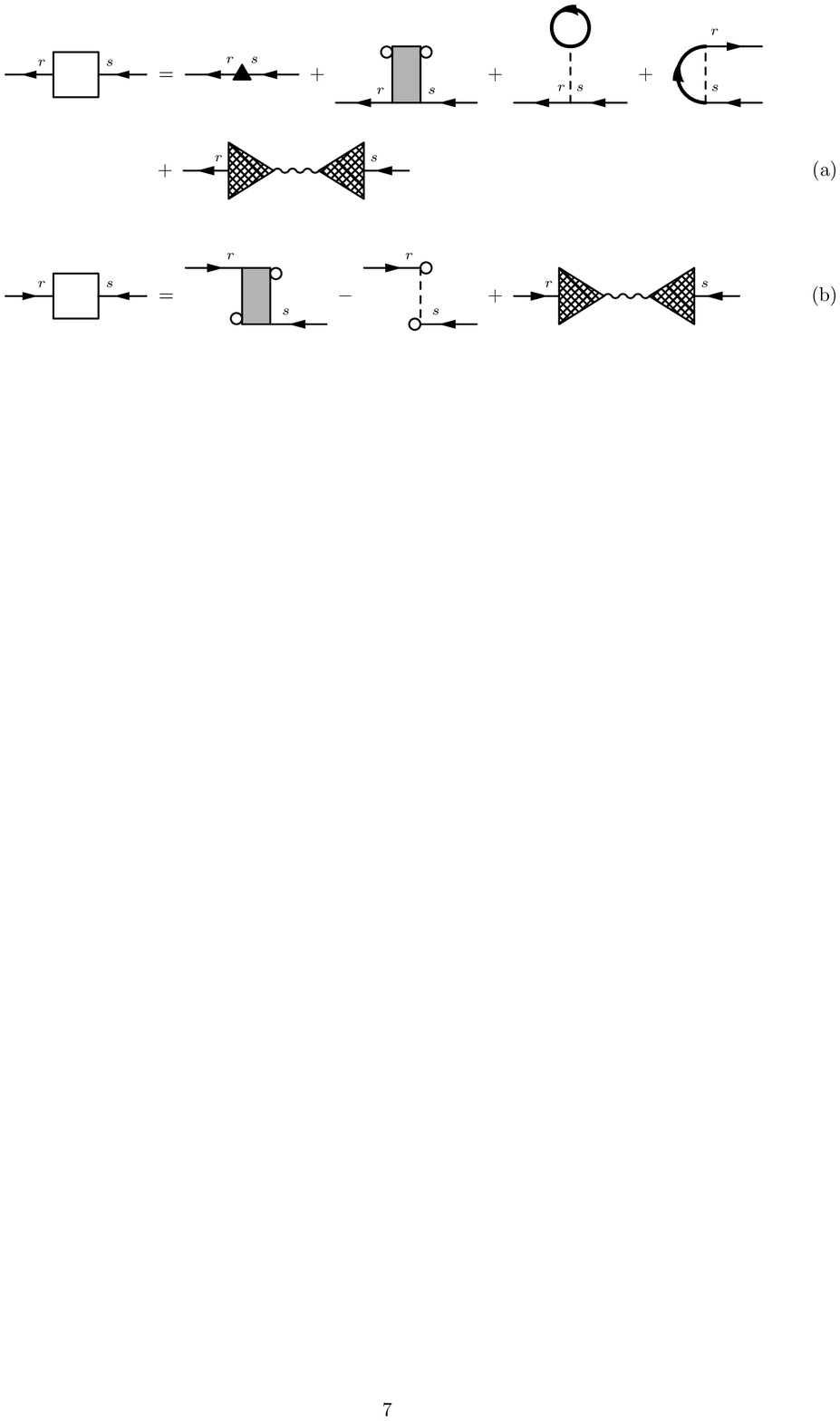}
  \caption{The Feynman diagrams contributing to the self-energies}
  \label{fig:nse1}
\end{figure}

Finally, for future use we remark that the Green's functions can be
categorized according to the amount of spin they transfer relative to
the spin of a condensate atom. For this aim we need the following
considerations. Equation \eqref{eq:dyson} splits into independent
parts according to the amount of spin transfered \cite{SzSz2}. For the
polar case, let us define the following matrices:
\begin{subequations}
  \label{eqs:gmatp}
  \begin{align}
    \gmat[0][\Gr]_{\gamma\delta}&=\left[\begin{array}{c c c}
        \Gr^{00}_{1,1} & \Gr^{00}_{1,-1}\\
        \Gr^{00}_{-1,1} & \Gr^{00}_{-1,-1}
      \end{array}\right]_{\gamma\delta}, &
    \gmat[0][\Sigma]_{\gamma\delta}&=\left[\begin{array}{c c c}
        \Sigma^{00}_{1,1} & \Sigma^{00}_{1,-1}\\
        \Sigma^{00}_{-1,1} & \Sigma^{00}_{-1,-1}
      \end{array}\right]_{\gamma\delta},\label{eq:gmat0p}\\
    \gmat[+][\Gr]_{\gamma\delta}&=\left[\begin{array}{c c c}
        \Gr^{--}_{1,1} & \Gr^{-+}_{1,-1}\\
        \Gr^{+-}_{-1,1} & \Gr^{++}_{-1,-1}
      \end{array}\right]_{\gamma\delta}, &
    \gmat[+][\Sigma]_{\gamma\delta}&=\left[\begin{array}{c c c}
        \Sigma^{--}_{1,1} & \Sigma^{-+}_{1,-1}\\
        \Sigma^{+-}_{-1,1} & \Sigma^{++}_{-1,-1}
      \end{array}\right]_{\gamma\delta},\label{eq:gmat+p}
  \end{align}
\end{subequations}
with the indices in the upper left corner corresponding to the amount
of spin transfered. For the ferromagnetic case we define
\begin{subequations}
  \label{eqs:gmatf}
  \begin{align}
    \gmat[0][\Gr]_{\gamma\delta}&=\left[\begin{array}{c c c}
        \Gr^{++}_{1,1} & \Gr^{++}_{1,-1}\\
        \Gr^{++}_{-1,1} & \Gr^{++}_{-1,-1}
      \end{array}\right]_{\gamma\delta}, &
    \gmat[0][\Sigma]_{\gamma\delta}&=\left[\begin{array}{c c c}
        \Sigma^{++}_{1,1} & \Sigma^{++}_{1,-1}\\
        \Sigma^{++}_{-1,1} & \Sigma^{++}_{-1,-1}
      \end{array}\right]_{\gamma\delta},\label{eq:gmat0f}\\
    \gmat[+][\Gr] & = \Gr^{00}_{1,1}, & \gmat[+][\Sigma]&= \Sigma^{00}_{1,1},
    \label{eq:gmat+f}\\
    \gmat[Q][\Gr] & = \Gr^{--}_{1,1}, & \gmat[Q][\Sigma]&= \Sigma^{--}_{1,1}.
    \label{eq:gmatQ+f}
  \end{align}
\end{subequations}
With the use of the above matrices, Eq. \eqref{eq:dyson} splits to
independent matrix equations, all of the form
\begin{equation}
  \label{eq:dysonmat}
  \gmat[n][\Gr]_{\alpha\gamma}(k)=\gmat[n][\Gr]_{(0)\alpha\gamma}(k)
  +  \gmat[n][\Gr]_{(0)\alpha\rho}(k)\gmat[n][\Sigma]_{\rho\sigma}(k)
  \gmat[n][\Gr]_{\sigma\gamma}(k),
\end{equation}
with $n$ being $0$, $+$ or $Q$. (Note however, that in the
ferromagnetic case for $n=+$ and $n=Q$ anomalous Green's functions do
not arise, i.e. $\alpha=\gamma=1$.)

\subsection{Collective excitations and generalized density correlation
  functions}

To describe collective excitations one can introduce the
correlation functions as follows \cite{SzSz2}:
\begin{subequations}
  \label{eqs:densc}
  \begin{align}
    &D_{nn}\ktau:=-\Big< T_\tau\big[ n\ktau n\adj(\vec{k}, 0)
    \big]\Big>\label{eq:d1},\\
    &D_{zz}\ktau:=-\Big< T_\tau\big[ {\mathcal{F}}_z\ktau
    {\mathcal{F}}_z\adj(\vec{k}, 0)\big]\Big>,\label{eq:d2}\\
    &D_{++}\ktau:=-\Big< T_\tau\big[ {\mathcal{F}}_+
    \ktau {\mathcal{F}}_+\adj(\vec{k},0)\big]\Big>,\label{eq:d3}\\
    &D^Q_{++}\ktau:=-\Big< T_\tau\big[ {\mathcal{F}}^Q_+
    \ktau {\mathcal{F}}^{Q\dagger}_+(\vec{k},0)\big]\Big>,\label{eq:d4}
  \end{align}
\end{subequations}  
for $k\neq0$, where the density operator:
\begin{subequations}
  \label{eqs:densops}
  \begin{equation}
    \label{eq:pardens}
    n(\vec{k})=\sum_{\vec{q}}a^\dagger_r(\vec{k}+\vec{q})a_r(\vec{q}),
  \end{equation}
  the spin density operator:
  \begin{equation}
    \label{eq:spindens}
    {\mathcal{F}}_z(\vec{k})=\sum_{\vec{q}}a^\dagger_r(\vec{k}+\vec{q})
    (F_z)_{rs}a_s(\vec{q}),
  \end{equation}
  a non self-adjoint operator generating spin waves:
  \begin{equation}
    \label{eq:swdens}
    {\mathcal{F}}_+(\vec{k})=\sum_{\vec{q}}a^\dagger_r(
    \vec{k}+\vec{q})(F_+)_{rs}a_s(\vec{q}),
  \end{equation}
  and another one for quadrupolar spin waves:
  \begin{equation}
    \label{eq:qswdens}
    {\mathcal{F}}^Q_+(\vec{k})=\sum_{\vec{q}}a^\dagger_r(
    \vec{k}+\vec{q})(F^2_+)_{rs}a_s(\vec{q}),\\
  \end{equation}
\end{subequations}
with $F_+=F_x + i F_y$. After Fourier transformations these
correlation functions can be expressed in the following way
\cite{SzSz2}:
\begin{subequations}
  \label{eqs:dpwm}
  \begin{align}
    D_{nn}(k) &= \left(\begin{array}{c} 1 \\ 1 \\ 1 \end{array}\right)
    \mat[0][D](k) \left(\begin{array}{c} 1 \\ 1 \\ 1 \end{array}\right),
    \label{eq:dpwm1}\\
    D_{zz}(k) &= \left(\begin{array}{c} 1 \\ 0 \\ -1 \end{array}\right)
    \mat[0][D](k) \left(\begin{array}{c} 1 \\ 0 \\ -1 \end{array}\right),
    \label{eq:dpwm2}\\
    D_{++}(k) &= 2 \left(\begin{array}{c} 1 \\ 1 \end{array}\right)
    \mat[+][D](k) \left(\begin{array}{c} 1 \\ 1 \end{array}\right),
    \label{eq:dpwm3}\\
    D^Q_{++}(k) &= 4 \mat[Q][D](k).\label{eq:dpwm4}
  \end{align}
\end{subequations}
The matrices on the right hand side satisfy the equations
\begin{equation}
  \label{eq:matpropeq}
  \mat[n][D](k)=\hslash \mat[n][\Pi](k) + \mat[n][\Pi](k)
  \mat[n][C] \mat[n][D](k),
\end{equation}
where $\mat[n][C]$ has been defined by Eqs. \eqref{eqs:strwc0} The polarization
function can be written as
\begin{equation}
  \label{eq:fsfpb}
  \mat[n][\Pi](k)=\mat[n][\Pi]^{(r)}(k) + \hslash^{-1}
  \gmat[n][\widetilde\Gr]_{\alpha\gamma}(k) \vv[][n][\widetilde\Lambda]_\alpha
  (k)\circ\vv[][n][\widetilde\Lambda]^\gamma(k).
\end{equation}
Here the first term on the right hand side represents the one-particle
line irreducible contributions, which in the Hartree-Fock
approximation are given by Eq. \eqref{eq:regpols}. Furthermore
$\widetilde\Lambda$ is defined in the Subsection \ref{sssec:avf} and
$\widetilde\Gr$ stands for the proper Green's function obtained as the
solution of Eq. \eqref{eq:dysonmat} with replacing $\Sigma$ by its
proper part (given in the Hartree-Fock approximation in Eqs.
\eqref{eq:pse1} and \eqref{eq:pse2}).

By introducing the so called dielectric function 
\begin{equation}
  \label{eq:dielmat}
  \mat[n][\eps](k)=\mat[n][1]-\mat[n][\Pi](k)\mat[n][C],
\end{equation}
the frequencies and the damping of collective excitations (after
analytic continuation to the second Riemann sheet) can be calculated
from the equation
\begin{equation}
  \label{eq:sptr0+}
  \det \mat[n][\eps](\vec{k},\omega)=0, 
\end{equation}
for $n=\{0,+\}$. For $n=Q$ they are given by the poles of $\mat[Q][\Pi]$,
since $\mat[Q][C]=0$.

It is important that the one-particle and collective spectra almost
coincide as can be proven by rearrangements of the contributions
\cite{SzSz2}. The spin density relaxation mode in the polar phase,
described by the correlation function \eqref{eq:dpwm2}, is an
exception, which does not have a counterpart among the one-particle
excitations. One can easily convince oneself that the general
treatment in \cite{SzSz2} directly applies to the Hartree-Fock model
worked out in the present paper.

\section{Applications}
\label{sec:appl}

In this section we evaluate the self-consistent Hartree--Fock
approximation, discussed in the previous section, to parameter values
relevant to possible experimental realizations. In most  cases we
use dimensionless quantities. Let us introduce the
following dimensionless small parameters:
\begin{subequations}
  \label{eqs:spar}
  \begin{align}
    \epsilon_n &:= \beta_0 N c_n,\\
    \epsilon_s &:= \beta_0 N |c_s|,\\
    \delta &:= \frac{|c_s|}{c_n}=\frac{\epsilon_s}{\epsilon_n},\label{eq:spar3}
  \end{align}
\end{subequations}
with $\beta_0:=1/(k_B T_0)$ the inverse transition temperature and $N$
the total particle number density. The transition temperature $T_0$ is
defined as the temperature, where the symmetric phase of the ideal
spin-1 Bose gas looses its stability, i.e. where
\begin{equation}
  \label{eq:ttemp}
  N=3 \frac{\Gamma\big(\frac{3}{2}\big) F\big(\frac{3}{2},0\big)}{4 \pi^2
    \lambda_0^3},
\end{equation}
with $\lambda_0:=\lambda(T=T_0)$ the thermal wavelength \eqref{eq:twl}
of the atom at temperature $T_0$. The self-energies are made
dimensionless with the help of the inverse temperature in the
following way:
\begin{equation}
  \label{eq:dlse}
  \sigma^{rs}_{\alpha\gamma}(k):=\beta\hslash\Sigma^{rs}_{\alpha\gamma}(k).
\end{equation}
The condensate density and the thermal densities in their
dimensionless form read as:
\begin{subequations}
  \label{eqs:dlde}
  \begin{align}
    \gamma_0&:=\beta c_n N_0,\\
    C_r&:=\beta c_n H^{rr}=\frac{\epsilon_n\sqrt{t}
      F\big(\frac{3}{2},\widehat\sigma^{rr}_{11}\big)}{3 \zeta\big(
      \frac{3}{2}\big)}, \label{eq:Crd}
  \end{align}
\end{subequations}
where $\beta=1/(k_B T)$ the inverse temperature. Automatic summation
is not applied to Eq. \eqref{eq:Crd}, the temperature is also used in
a dimensionless form, with $t=T/T_0$ and $\zeta(s)$ is the Riemann
zeta function. The last equation is obtained with the use of Eq.
\eqref{eq:hartree}. The dependence on the wavenumber is also evaluated
in a dimensionless form. For this aim the thermal wavelength is used
again
\begin{equation}
  \label{eq:wnd}
  u:=|\vec{k}|\lambda.
\end{equation}

One has to recall at this place that the Hartree-Fock approximation is
not applicable in the vicinity of the phase transition. Furthermore,
it is not valid at very low temperatures where the contributions of
Beliaev processes might become important like in the case of a scalar
condensate \cite{FS1}. Nonetheless there is a considerable
temperature range where one can expect that such a Hartree-Fock
approach gives quite accurate results.

\subsection{Ferromagnetic system}
\label{ssec:ferr}

This system is realized by the gas of $^{87}\mathrm{Rb}$ atoms, which
has been recently trapped and condensed in an all optical way
\cite{BSC}. As stated earlier, the condensate spinor is chosen to be
$\zeta_r=(1,0,0)^T$. We consider a unit volume system with transition
temperature at $T_0=200\;\mathrm{nK}$ (close to that analytically
calculated in Ref.  \cite{HGT} for the experimental situation realized
in Ref.  \cite{BSC}), which by Eq.  \eqref{eq:ttemp} yields a
homogeneous total density $N=10^{20}\;\mathrm{m^{-3}}$. We take the
scattering lengths as $a_0=104a_\textrm{B}$, and
$a_2=102a_\textrm{B}$ ( with $a_\textrm{B}$ being the Bohr-radius)
according to the calculations made in Ref.  \cite{KBG}( note that
their results have some uncertainty, since the exact number of bound
states in the triplet potential of two $^{87}\mathrm{Rb}$ atoms is not
known exactly).  According to Eqs.  \eqref{eqs:spar}, the values of
the dimensionless small parameters are: $\epsilon_n=0.2$ and
$\delta=0.0065$.

The length of the condensate is the longest length scale of the
system.  Taking the length of
the condensate as $100\,\mathrm{\mu m}$ means a minimal wavenumber
$k_{\mathrm{min}}=10^4\,\mathrm{m^{-1}}$. For experimental
applications local excitations should be considered. One can take the
length scale of local perturbations to $10\,\mathrm{\mu m}$ resulting
in a typical wavenumber $k_{\mathrm{typ}}=10^{5}\,\mathrm{m^{-1}}$.

With the use of Eqs. \eqref{eq:intse}, \eqref{eq:spar3},
\eqref{eq:dlse}, \eqref{eqs:dlde} and \eqref{eq:fchp} the
dimensionless self-energies of the internal lines read as:
\begin{subequations}
  \label{eqs:dlsef}
  \begin{align}
    \widehat\sigma^{++}_{11} &= (1-\delta)\gamma_0,\\
    \widehat\sigma^{00}_{11} &= -(1-\delta)C_+ +
    (1+\delta) C_0 - 2 \delta C_-,\\
    \widehat\sigma^{--}_{11} &= -(1-3\delta)C_+ +
    (1-3\delta)C_-+2\delta\gamma_0.\\
    \intertext{Expressing the condensate density as the total density
      minus the noncondensate densities leads to the equation:}
    \gamma_0&=\frac{\epsilon_n}{t}-C_+-C_0-C_-.\label{eq:tdsncd}
  \end{align}
\end{subequations}
Equations \eqref{eqs:dlsef} and \eqref{eq:Crd} form a closed set of
nonlinear equations for the densities and self-energies 
, which can be
solved numerically.

\begin{figure}[!ht]
  \centering
  \includegraphics*[32mm,226mm][96mm,279mm]{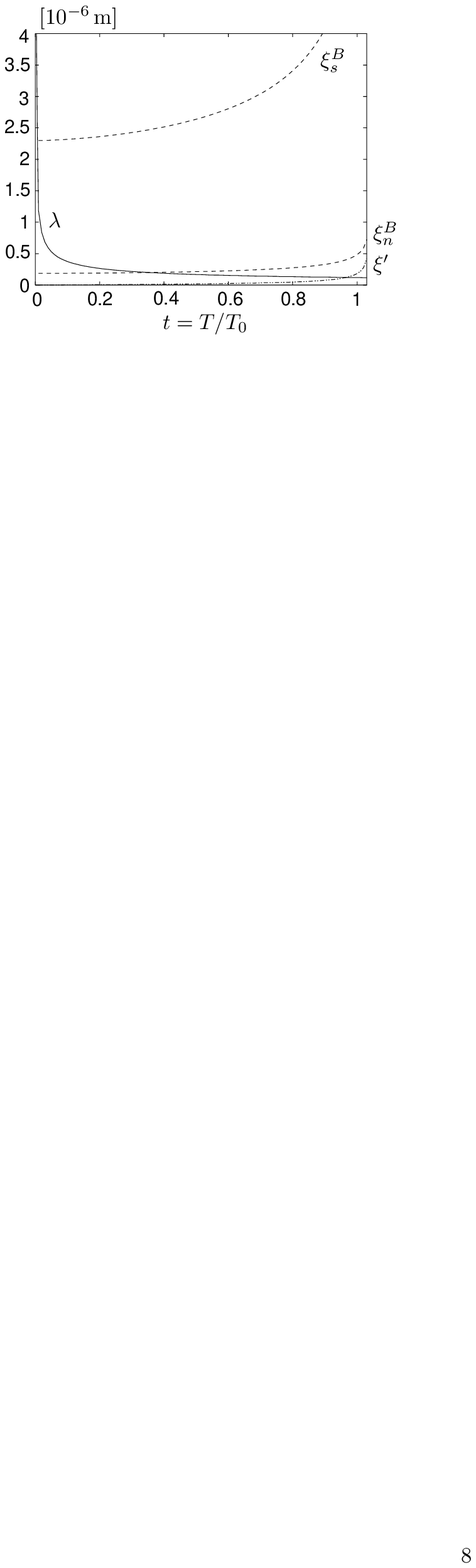}
  \caption{The characteristic lengths of the ferromagnetic $^{87}\mathrm{Rb}$
    system as functions of the dimensionless temperature $t=T/T_0$.}
  \label{fig:clf}
\end{figure}

\subsubsection{Zero spin transfer excitations}
\begin{figure}[!ht]
  \centering
  \includegraphics*[32mm,228mm][160mm,279mm]{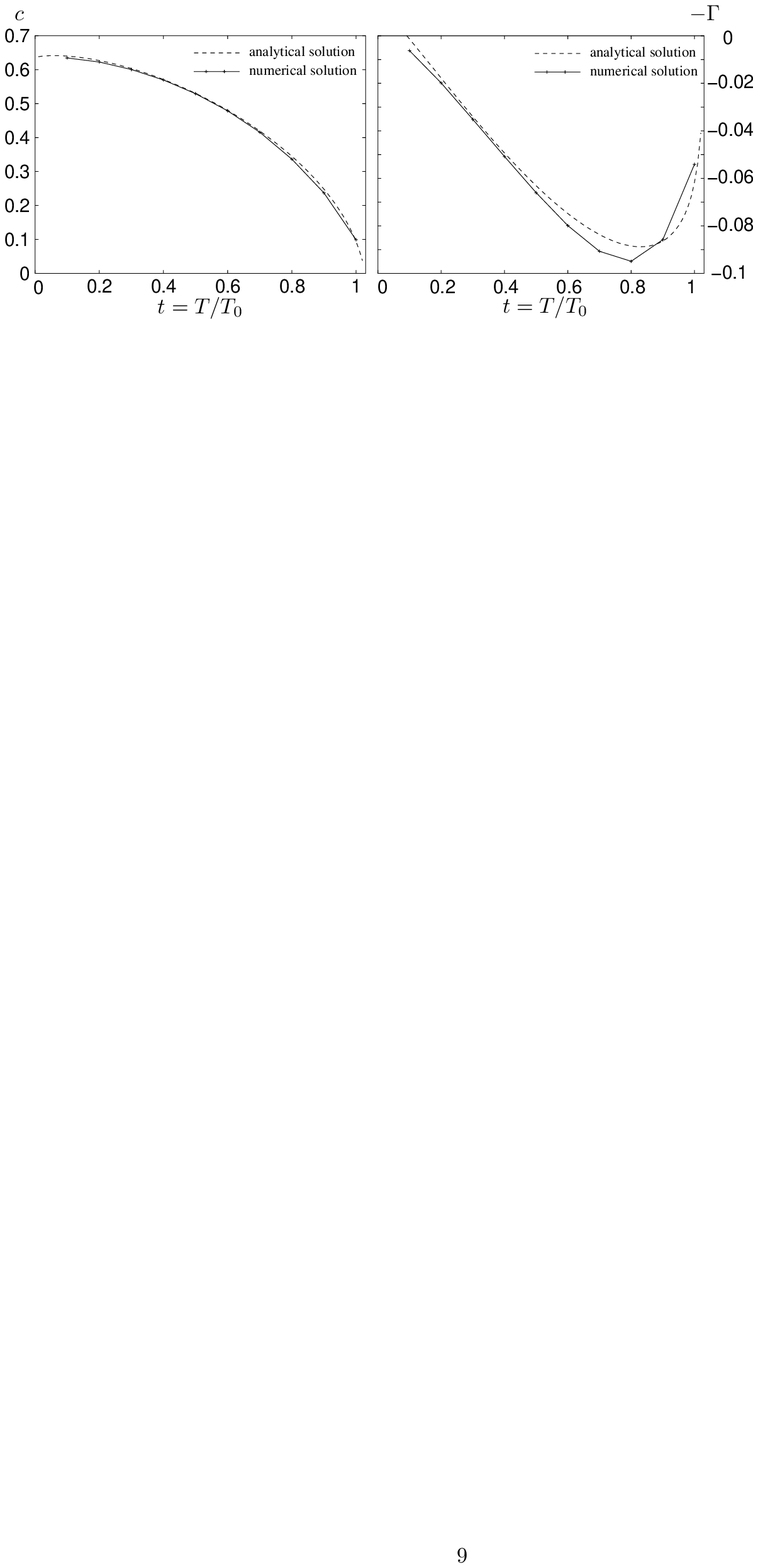}
  \caption{The sound velocity $c$, and the parameter $-\Gamma$ of the
    $^{87}\mathrm{Rb}$ ferromagnetic system in units of $k_B T_0
    \lambda_0/\hslash$, and as the function of $t=T/T_0$.}
  \label{fig:cefpd}
\end{figure}

Equations \eqref{eq:dpwm1} and \eqref{eq:dpwm2} show, that density and
spin density collective excitations fall to this category together
with the one-particle excitations determined by  the zero 
spin transfer Green's
function Eq. \eqref{eq:gmat0f}. According to Eq. \eqref{eq:thetasol}
the matrix $\mat[n][\Theta](k_1,k_2,q)$ does not depend on $q$,
moreover it depends only on $k_2-k_1$. Using this property and Eqs.
\eqref{eq:pse1}, \eqref{eq:pse2}, \eqref{eqs:dlde} and \eqref{eq:fchp}, the
dimensionless proper self-energy of spin-transfer $0$ can be brought
to the form:
\begin{equation}
  \label{eq:dse0f}
  \gmat[0][\widetilde\sigma]_{\alpha\gamma}(k)=\beta N_0\big[
  \Theta^{++}_{++}(0,k,0)-(c_n+c_s)\big],
\end{equation}
for all $\alpha,\gamma$. The zeroth order contributions of the
anomalous vertex vectors \eqref{eq:vectvert0fer} read as:
\begin{equation}
  \label{eq:vv0zf}
  \vv[][0][\Lambda]_{(0),\alpha}=\sqrt{N_0}\left(\begin{array}{c}
      1\\0\\0\end{array}\right).
\end{equation}
With the help of Eqs. \eqref{eq:thetasol}, \eqref{eq:dielstar},
\eqref{eq:regpols}, \eqref{eq:pav1} and \eqref{eq:pav2} the regular
polarization, the anomalous vertex and the proper self-energy can be
given explicitly. With the use of these and Eqs. \eqref{eq:fsfpb} and
\eqref{eq:dielmat} the dielectric function can be expressed
explicitly. The calculation is straightforward but rather lengthy.
According to Eq. \eqref{eq:sptr0+} one way to determine the spectrum of
density and spin-density excitations is provided by the equation:
\begin{equation}
  \label{eq:de0zf}
  \det\mat[0][\eps](\vec{k},\omega)=0.
\end{equation}
Equation \eqref{eq:de0zf} can be directly solved numerically.
Analytically solvable approximations can also be obtained by keeping
the regular terms and the contribution of the nearest pole to the real
axis from the Mittag--Leffler representation of the spectral function
of the bubble graphs occurring \cite{SzSz2}.  This approximation is
valid for a linear excitational spectrum when the mean-field
characteristic lengths are the largest length scales in the system,
i.e. when $\xi^B_{n,s}\gg\lambda, \xi'$.  Here $\xi^{B}_{n,s} =
\hslash / \sqrt{4MN_0|c_{n,s}|}$ and $\xi' = M / (4 \pi \hslash^2 N_0
\beta)$.  The characteristic lengths of the system are depicted in
Fig.  \ref{fig:clf}. It is visible that the analytical approximation
can be used for $t>0.35$.  Both the real and imaginary part of the
solution of Eq.  \eqref{eq:de0zf} is found to depend linearly on
momentum for the relevant range. the solution can be written in the
form:
\begin{equation}
  \label{eq:dwf}
  \omega = c \; |\vec{k}| - i \Gamma \; |\vec{k}|,
\end{equation}
where $c$ is the speed of the excitation and $\Gamma$ is also a
positive constant yielding the damping. The numerical and analytical
solutions are plotted in Fig. \ref{fig:cefpd}.  These kind of
excitations (density and spin density excitations) show similar
features than the density excitations of the scalar gas \cite{FRSzG}.
They have linear dispersion curve for small wave number with an also
linear and small imaginary part.

\subsubsection{Plus two spin transfer excitations}
\label{sssec:pl2f}

As indicated by Eq. \eqref{eq:dpwm4} quadrupolar spin waves fall into
this category together with the one particle excitations given by the
$+2$ spin transfer Green's function Eq. \eqref{eq:gmatQ+f}. Spin
conservation forbids here the existence of anomalous Green's
functions.  Furthermore since $^QC=0$ both the Green's function
$^Q\Gr$ and the correlation function $^QD$ are proper. Using Eqs.
\eqref{eq:pse1}, \eqref{eqs:dlde} and \eqref{eq:fchp}, the
dimensionless proper self-energy of spin-transfer $2$ can be brought
to the form:
\begin{equation}
  \label{eq:dse+qf}
  \gmat[Q][\widetilde\sigma]_{11}(k)=\beta N_0\big[
  \Theta^{++}_{--}(0,k,0)-(c_n+c_s)\big]
  + (1-3 \delta)(C_- - C_+).
\end{equation}
With the help of Eqs.
\eqref{eqs:thetaQ}, \eqref{eq:thetasol} and \eqref{eq:dielstar} the
$\Theta$ matrix for $n=Q$ can be cast to the form:
\begin{equation}
  \label{eq:thetaQf}
  \Theta^{++}_{--}(0,k,0)=\frac{c_n(1+\delta)}{1-c_n(1+\delta)
    \Pi^{-+}_{(0)+-}(k)}.
\end{equation}
According to Eqs. \eqref{eq:dysonmat}, \eqref{eq:dlse},
\eqref{eq:dse+qf} and \eqref{eq:thetaQf} the Green's function
$\gmat[Q][\Gr]$ is expressed (after the usual analytical continuation
in the frequency) as
\begin{equation}
  \label{eq:grqf}
  \gmat[Q][\Gr](\vec{k},\omega)=\gmat[Q][\widetilde\Gr](\vec{k},\omega)
  =\frac{\beta \hslash}{u^2(\Omega-1)-\gmat[Q][\widetilde\sigma]_{11}
    (\vec{k},\omega)},
\end{equation}
with $u$ defined by Eq. \eqref{eq:wnd}, and
$\Omega:=\hslash\omega/\kink$. The poles of the Green's function
\eqref{eq:grqf} with the use of Eqs.  \eqref{eq:dse+qf},
\eqref{eq:thetaQf} and \eqref{eqs:dlde} are determined by the
following equation:
\begin{widetext}
  \begin{equation}
    \label{eq:exceqf}
    u^2(\Omega-1)+(1-3\delta)(C_+-C_-)-2\delta\gamma_0
    -(1+\delta)c_n\Pi^{-+}_{(0)+-}(u,\Omega)\big[u^2(\Omega-1)+(1-3\delta)
    (C_+-C_-)+\gamma_0(1-\delta)\big] = 0.
  \end{equation}
\end{widetext}
Equation \eqref{eq:exceqf} can be solved  in a numerical way
for $\Omega$ as a function of $u$.  Analytical approximate solution
is also possible with the help of Eqs. \eqref{eqs:spfd} and
\eqref{eq:spfuncdf}. If one is also interested in the damping rates,
the small imaginary part \eqref{eq:spfd2} of the bubble graph is also
to be considered. The following form is motivated in Appendix
\ref{sec:absa}:
\begin{equation}
  \label{eq:bubasim}
  c_n \Im \; \Pi^{-+}_{(0)+-}(u,\Omega)=e^{-\left(\frac{\Delta
        \widehat\sigma}{2 u}\right)^2}(\ldots), \text{ for } \Im\,\Omega=0,
\end{equation}
with $\Delta \widehat \sigma=\widehat \sigma^{--}_{11} - \widehat
\sigma^{++}_{11}$. (We have supposed that the imaginary part of
$\Omega$ is very small, therefore $\Im \;\Pi=\Pi''$.) The quantity in
the brackets is not important for our  considerations here.

\begin{figure}[!ht]
  \centering
  \includegraphics*[32mm,226mm][162mm,280mm]{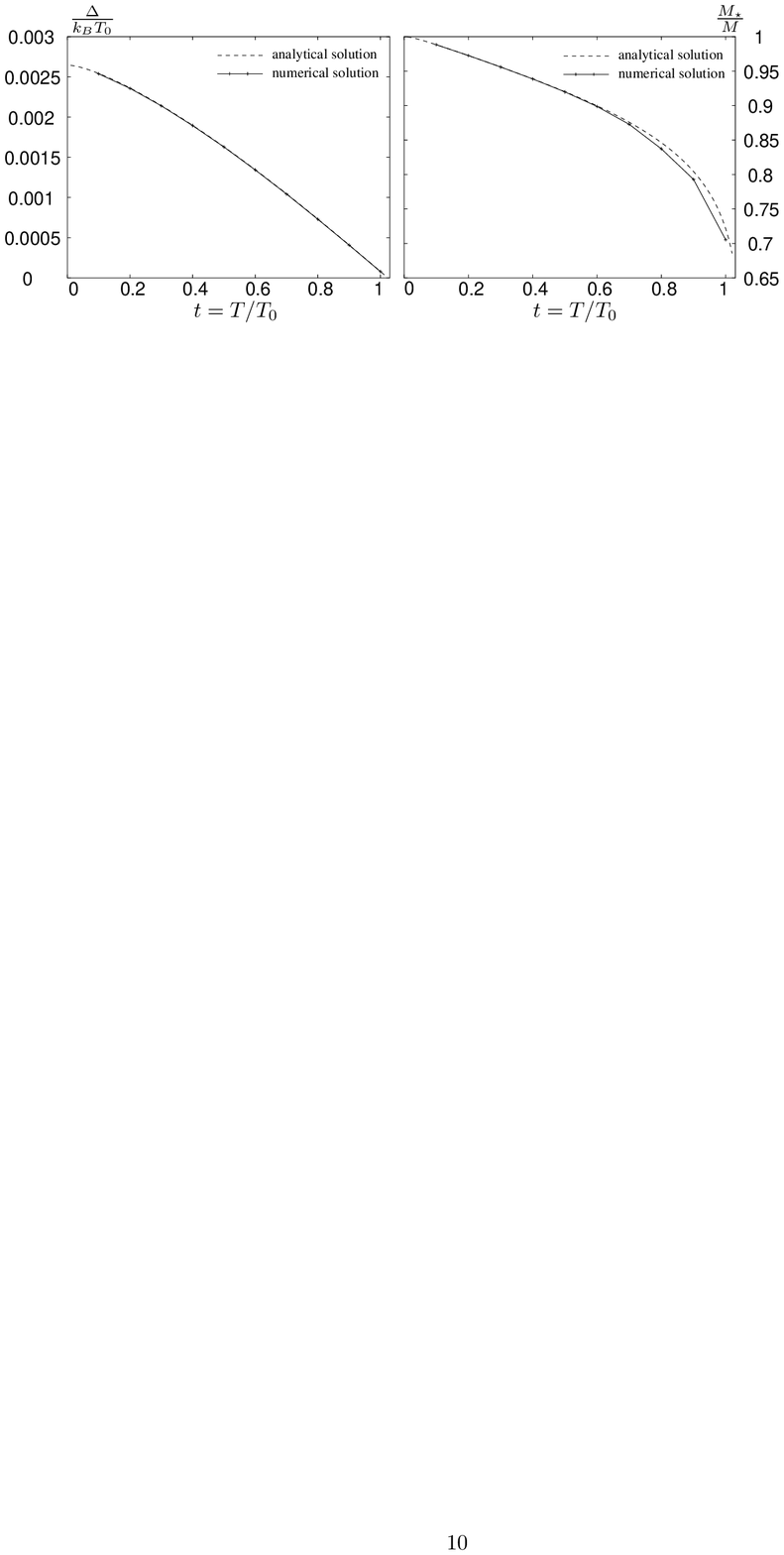}
  \caption{The gap $\Delta$ and effective mass of the quadrupolar spin wave
    excitations as functions of the temperature $t=T/T_0$. The gap is
    measured in units of $k_B T_0$, while $M_\star$ is measured in the
    units of the bare mass of the $^{87}\mathrm{Rb}$ atom.}
  \label{fig:gemf}
\end{figure}
\begin{figure}[!ht]
  \centering
  \includegraphics*[32mm,226mm][108mm,279mm]{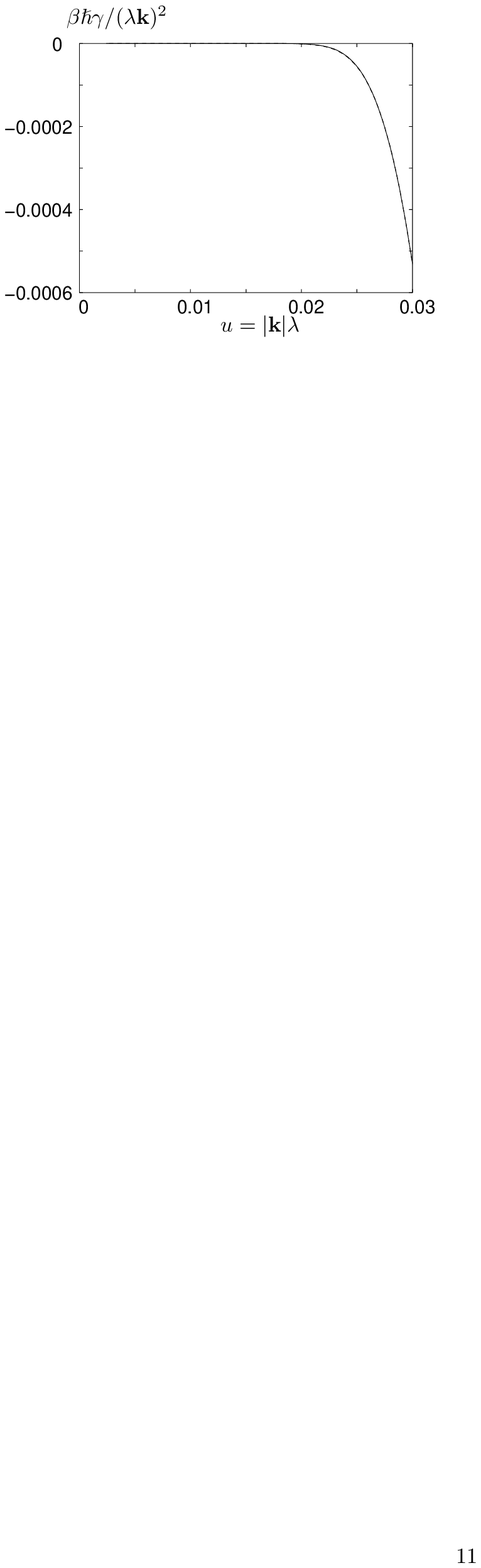}
  \caption{The damping rate of the quadrupolar spin wave excitations at
    temperature $T=0.7\,T_0$. The figure contains the data points of
    the numerical solution and the curve of the fitting function
    \eqref{eq:fit}.}
  \label{fig:dqf}
\end{figure}

The long wavelength excitations can be written in the form:
\begin{equation}
  \label{eq:qswef}
  \omega=\frac{\Delta}{\hslash}+\frac{\hslash \vec{k}^2}{2 M_\star} +
  i \gamma(|\vec{k}|),
\end{equation}
with $\Delta$ being the energy gap of the quadrupolar excitations,
$M_\star$ their effective mass and $\gamma(|\vec{k}|)$ is their
wavenumber dependent damping rate. The gap $\Delta$ and the effective
mass $M_\star$ is plotted in Fig. \ref{fig:gemf}. The imaginary part
of the numerical solution  at $T=0.7\,T_0$
temperature is plotted in Fig. \ref{fig:dqf}.  According to Eq.
\eqref{eq:bubasim} and the numerical solution of Eqs.
\eqref{eqs:dlsef}, $\gamma(|\vec{k}|)$ vanishes exponentially for
$|\vec{k}|\rightarrow0$. This exponential suppression 
dominates below the
characteristic value:
\begin{equation}
  \label{eq:cutmom}
u_c=\frac{|\widehat\sigma^{--}_{11} - \widehat\sigma^{++}_{11}|}{2}
\approx0.073.  
\end{equation}
For illustration we have fitted the function
\begin{equation}
  \label{eq:fit}
  f(u)=- C e^{\left(\frac{u_c}{u}\right)^2}
\end{equation}
to the numerical values in Fig. \ref{fig:dqf}, and found
$C\approx0.089$ and $u_c\approx0.068$ in good agreement with
\eqref{eq:cutmom}.

\subsubsection{Plus one spin transfer excitations}

\begin{figure}[!ht]
  \centering
  \includegraphics*[32mm,223mm][100mm,279mm]{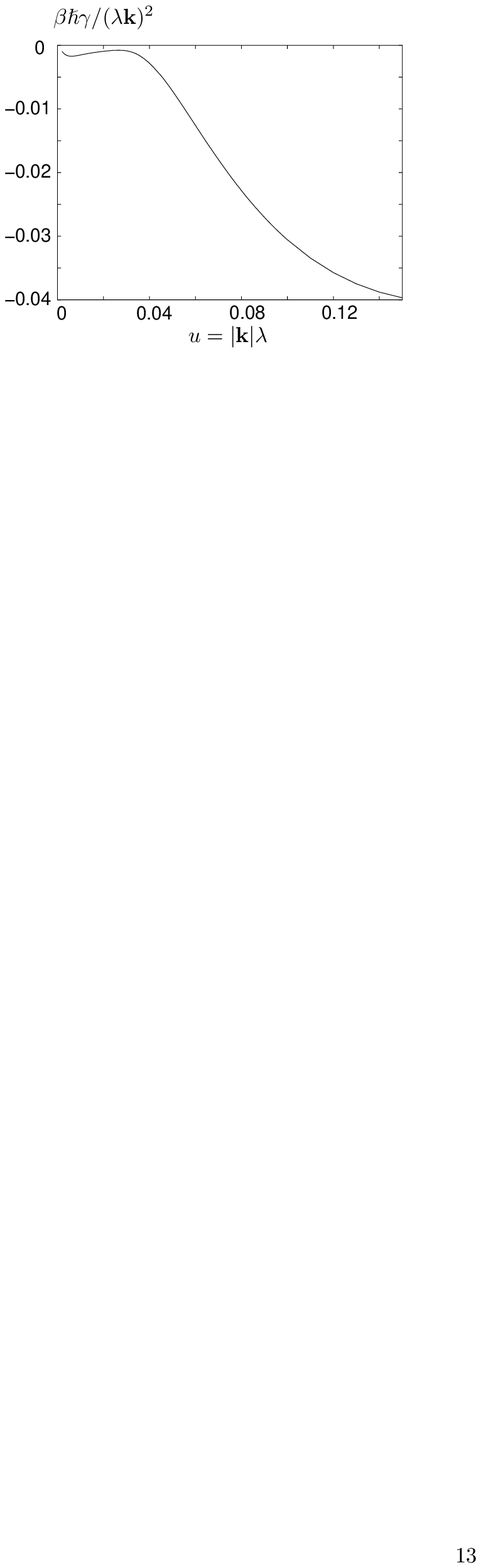}
  \caption{The damping rate divided by $u^2$ of the spin wave excitations in
    units of $k_B T/\hslash$ at $T= 0.8\, T_0$.}
  \label{fig:swdf}
\end{figure}

As indicated by Eq. \eqref{eq:dpwm3} spin waves fall into this
category together with the one particle excitations determined by the
$+1$ spin transfer Green's function Eq. \eqref{eq:gmat+f}. Spin
conservation forbids here the existence of anomalous Green's
functions. Using Eqs.  \eqref{eq:pse1}, \eqref{eqs:dlde} and
\eqref{eq:fchp}, the dimensionless proper self-energy of spin-transfer
$1$ can be brought to the form:
\begin{equation}
  \label{eq:dse+f}
  \gmat[+][\widetilde\sigma]_{11}(k)=\beta N_0\big[
  \Theta^{++}_{00}(0,k,0)-(c_n+c_s)\big]
  -(1-\delta)C_+ + (1+\delta)C_0 - 2
  \delta C_-.
\end{equation}
The zeroth order contribution for the anomalous vertex vectors Eq.
\eqref{eq:vectvert+fer} are
\begin{equation}
  \label{eq:vv+zf}
  \vv[][+][\Lambda]_{(0),1}=\sqrt{N_0}\left(\begin{array}{c}
      1\\0\end{array}\right),\qquad \vv[][+][\Lambda]_{(0),-1}=\vv[][][0].
\end{equation}
Again with the help of Eqs. \eqref{eq:thetasol}, \eqref{eq:dielstar},
\eqref{eq:regpols}, \eqref{eq:pav1} and \eqref{eq:pav2} the regular
polarization, the anomalous vertex and the proper self-energy can be
given explicitly. With the use of these and Eqs. \eqref{eq:fsfpb} and
\eqref{eq:dielmat} the dielectric function can be expressed. The
spectrum of the spin wave excitations is given as the solution of the
equation:
\begin{equation}
  \label{eq:de+zf}
  \det\mat[+][\eps](\vec{k},\omega)=0,
\end{equation}
\begin{figure}[!ht]
  \centering
  \includegraphics*[32mm,226mm][98mm,280mm]{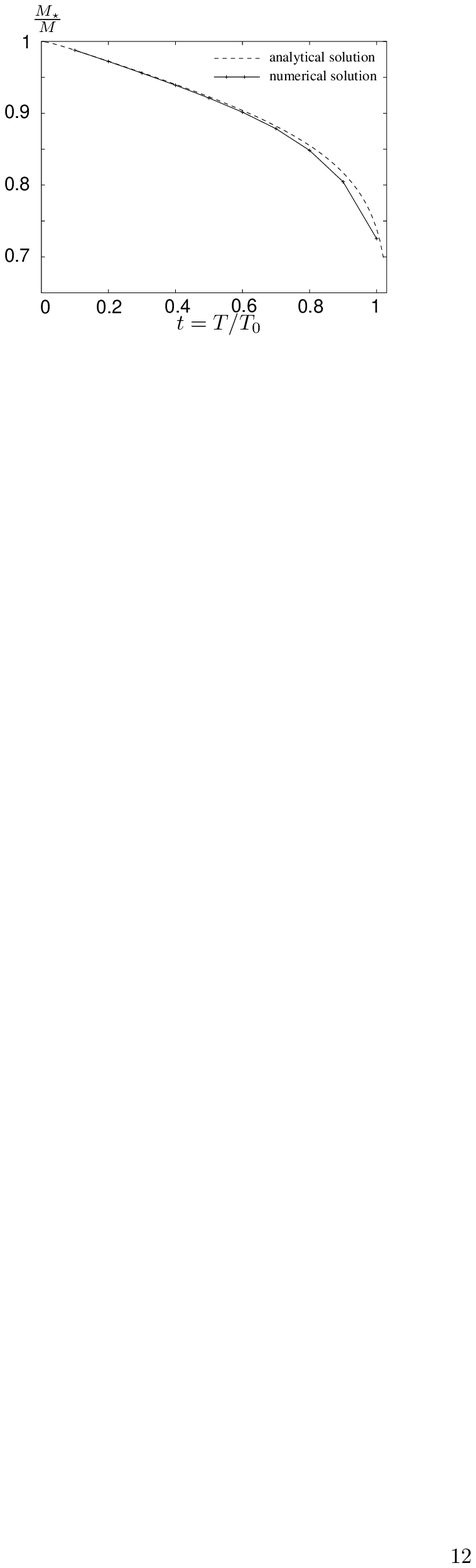}
  \caption{The effective mass of the spin wave excitations in units of the
    bare mass of the $^{87}\mathrm{Rb}$ atom as a function of
    $t=T/T_0$.}
  \label{fig:swemf}
\end{figure}
The energy of the excitations is found to be quadratic for small
wavenumber (including wavenumber values considered in this paper). The
spectrum of excitations can be written in the form:
\begin{equation}
  \label{eq:swf1}
  \omega = \frac{\hslash \vec{k}^2}{2 M_\star} + i \gamma(|\vec{k}|),
\end{equation}
with $M_\star$ the effective mass and $\gamma(|\vec{k}|)$ the damping
rate.  The temperature dependence of the effective mass is plotted in
Fig. \ref{fig:swemf}, while the damping rate is plotted for
$T=0.8\,T_0$ in Fig. \ref{fig:swdf}.
The behavior of the damping rate of the spin wave excitations is more
complex than that of the quadrupolar spin waves shown in Fig.
\ref{fig:dqf}. The reason is that in this case two different bubble
graphs emerge. One is $\Pi^{0+}_{+0}$ with self-energy difference
$\Delta \widehat \sigma_1=\widehat \sigma^{00}_{11}-\widehat \sigma^{++}_{11}$,
the other is $\Pi^{-0}_{0-}$ with self-energy difference $\Delta \widehat
\sigma_2=\widehat \sigma^{--}_{11}-\widehat \sigma^{00}_{11}$. As a
consequence there are two characteristic wavenumber values, where the
damping rates suffer a drastic change. At $t=0.8$ these dimensionless
wavenumbers are:
\begin{subequations}
  \label{eqs:dswcuf}
  \begin{align}
    u_{c,1}&=\frac{|\widehat \sigma^{--}_{11}-\widehat \sigma^{00}_{11}|}{2}
    \approx 0.0005,\\
    u_{c,2}&=\frac{|\widehat \sigma^{00}_{11}-\widehat \sigma^{++}_{11}|}{2}
    \approx 0.05.
  \end{align}
\end{subequations}
Since the first wavenumber value is much smaller than the minimal
accessible wavenumber $u_{c,1}\ll \lambda k_{\mathrm{min}}$, there is some
damping of the long wavelength spin wave excitations even, when
quadrupolar spin waves are not damped.

\subsection{Polar system}
\label{ssec:polar}

This system is realized by the ultracold gas of $^{23}\mathrm{Na}$
atoms, when the magnetic field is sufficiently small \cite{SKea,Sea1}.
We consider a pure condensate and take its spinor as $\zeta_r =
(0,1,0)^T$. The system considered is also of unit volume with periodic
boundary conditions and a transition temperature of $T_0 =
2\,\mathrm{\mu K}$. The total particle density calculated with Eq.
\eqref{eq:ttemp} is $N=4.6\c 10^{20}\mathrm{m}^{-3}$. We take the
scattering lengths of the sodium atoms from Ref. \cite{Crea} to be
$a_0 = 50 a_B$ and $a_2 = 55 a_B$. According to Eqs. \eqref{eqs:spar},
the values of the dimensionless small parameters are:
$\epsilon_n=0.17$ and $\delta=0.031$. To determine the interesting
wavenumber range, we also take the condensate size to
$100\,\mathrm{\mu m}$, from which the minimal wavenumber is
$k_{\mathrm{min}} = 10^4 \mathrm{m}^{-1}$. The length scale of local
perturbations is chosen to correspond to a typical
wavenumber value $k_{\mathrm{typ}}=10^5\,\mathrm{m}^{-1}$.

With the use of Eqs. \eqref{eq:intse}, \eqref{eq:dlse},
\eqref{eqs:dlde} and Eq. \eqref{eq:pchp}, and the $(+,-)$ symmetry of
the polar case [namely that $\widehat \sigma^{--}_{11}=\widehat
\sigma^{++}_{11}$], the dimensionless self-energies of the internal
lines read as:
\begin{subequations}
  \label{eqs:dlsep}
  \begin{align}
    \widehat\sigma^{00}_{11} &= \gamma_0,\\
    \widehat\sigma^{++}_{11} &= (1-\delta)(C_+ - C_0) +\delta\gamma_0.\\
    \intertext{Expressing again the condensate density as the total
      density minus the noncondensate densities and using the 
      symmetry mentioned leads to the final equation:}
    \gamma_0&=\frac{\epsilon_n}{t}-2C_+-C_0.\label{eq:tdsncd2}
  \end{align}
\end{subequations}
Equations \eqref{eqs:dlsep} and \eqref{eq:Crd} are closed and
numerically solvable for the densities and self-energies in the polar
 phase.

\begin{figure}[!ht]
  \centering
  \includegraphics*[33mm,225mm][97mm,279mm]{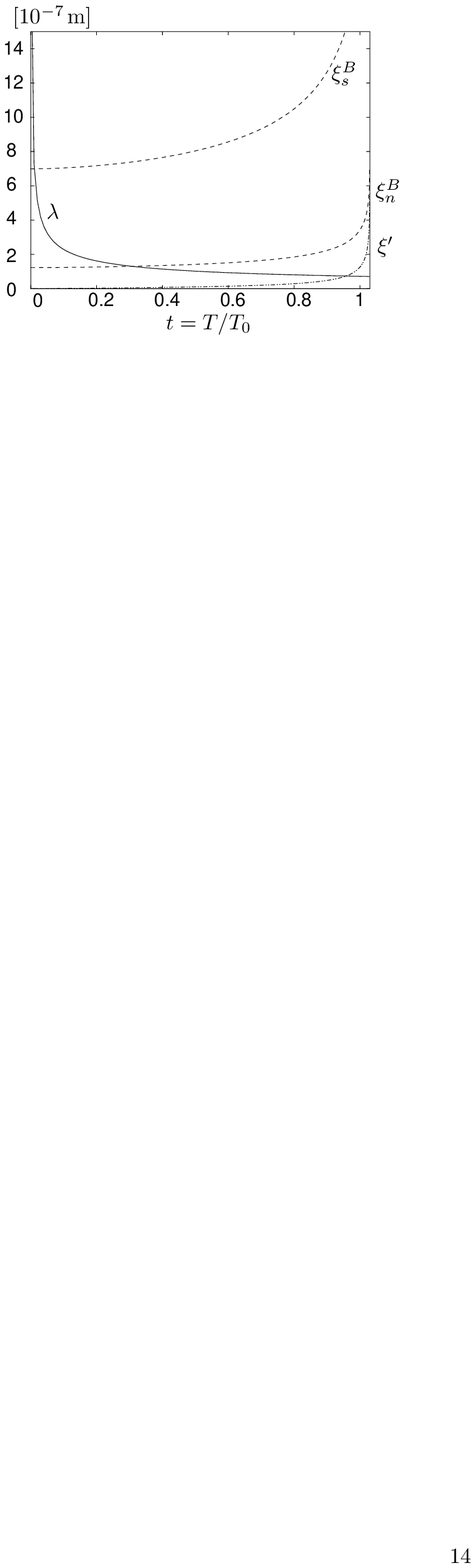}
  \caption{The characteristic lengths of the polar $^{23}\mathrm{Na}$ system
    as a function of temperature $t=T/T_0$.}
  \label{fig:scf}
\end{figure}

The characteristic lengths of the polar system are plotted in Fig.
\ref{fig:scf}. As can be seen from the figure, the approximation is
usable roughly for $t>0.3$ (when the largest characteristic lengths
are the mean field correlation lengths).

\subsubsection{Zero spin transfer excitations}

As in the case of the ferromagnetic system, these type of excitations 
are the
particle density and spin density excitations defined as the poles of
the correlation functions \eqref{eq:d1} and \eqref{eq:d2} [See Eqs.
\eqref{eq:dpwm1} and \eqref{eq:dpwm2}]. Since the $\Theta$ matrix does
not depend on the momentum transferred [See the remark after Eqs.
\eqref{eqs:thetasols}], the proper self-energy of spin-transfer 0 for
the polar system can be cast to the form (by using Eqs.
\eqref{eq:pse1}, \eqref{eq:pse2}, \eqref{eqs:dlde} and
\eqref{eq:pchp}):
\begin{equation}
  \label{eq:dse0p}
  \gmat[0][\widetilde\sigma]_{\alpha\gamma}(k)=\beta N_0\big[
  \Theta^{00}_{00}(0,k,0)-c_n\big],
\end{equation}
for all $\alpha,\gamma$. The zeroth order contributions of the
anomalous vertex vectors \eqref{eq:vectvert0pol} read as:
\begin{equation}
  \label{eq:vv0zp}
  \vv[][0][\Lambda]_{(0),\alpha}=\sqrt{N_0}\left(\begin{array}{c}
      0\\1\\0\end{array}\right).
\end{equation}
With a lengthy but trivial calculation, using Eqs.
\eqref{eq:thetasol}, \eqref{eq:dielstar}, \eqref{eq:regpols},
\eqref{eq:pav1}, \eqref{eq:pav2},\eqref{eq:fsfpb} and
\eqref{eq:dielmat}, one can give the explicite 
form of the dielectric
function $\mat[0][\eps]\komega$.  Analytically continuing it in the
usual way in the $\omega$ variable one determines the elementary
excitations by 
\begin{equation}
  \label{eq:de0zp}
  \det\mat[0][\eps](\vec{k},\omega)=0.
\end{equation}
It can be seen from the detailed  calculation, that the
left hand side of Eq. \eqref{eq:de0zp} for the polar case is the
product of two terms. One term determines the particle density
excitations and the other term determines the spin density
excitations. This behavior is not limited to this approximation. As
shown generally in Ref. \cite{SzSz2}, the particle density and spin
density correlation functions  $D_{nn}$ and $D_{zz}$, 
respectively, have
different excitation spectra in the polar case, even if they both
belong to spin transfer zero.

\begin{figure}[!ht]
  \centering
  \includegraphics*[33mm,225mm][97mm,279mm]{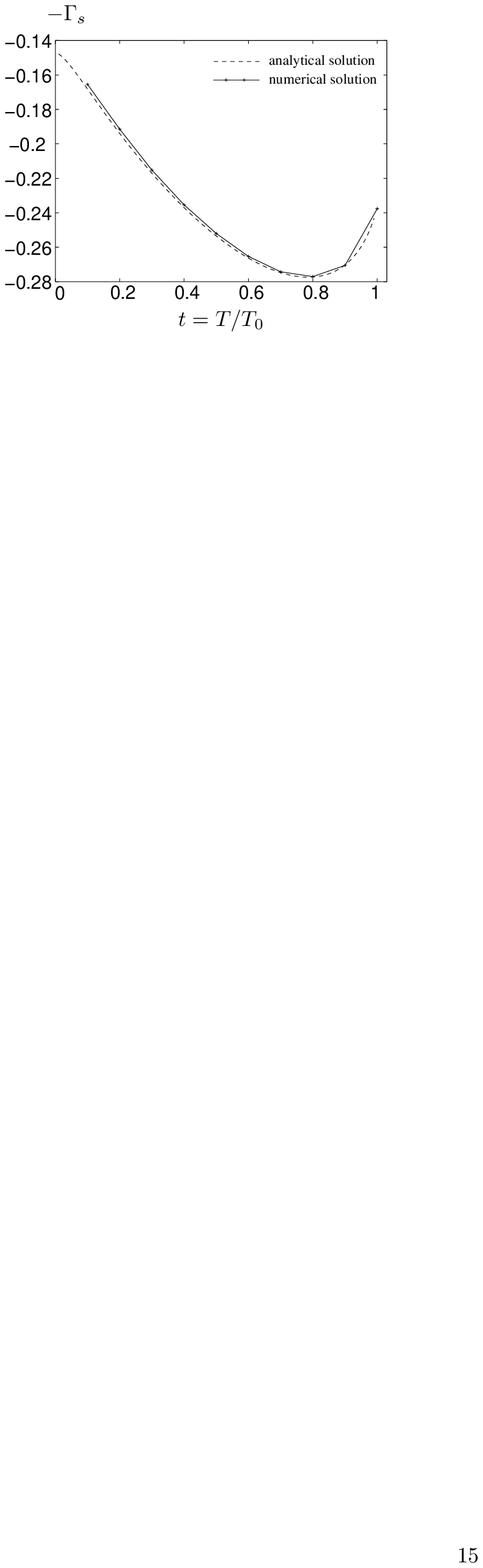}
  \caption{The relaxation rate parameter $-\Gamma_s$ of the spin density wave
    of the $^{23}\mathrm{Na}$ (polar) system in units of $k_B T_0
    \lambda_0/\hslash$, and as the function of $t=T/T_0$.}
  \label{fig:sdip}
\end{figure}

The spin density excitations are the solutions of the following
equation:
\begin{equation}
  \label{eq:sdep}
  1-(1+3\delta)c_n\Pi^{++}_{(0)++}(u,\Omega) = 0,
\end{equation}
with $\Omega=\hslash \omega /\kink$ and $u=|\vec{k}|\lambda$. This is an
overdamped mode, its frequency is purely imaginary with
linear wavenumber dependence:
\begin{equation}
  \label{eq:sdwcp}
  \omega= -i\Gamma_s \,|\vec{k}|,
\end{equation}
where $\Gamma_s$ is a positive parameter with the dimensionality of
velocity. The numerical and analytical solution is plotted in Fig.
\ref{fig:sdip}. The analytical solution is obtained from the
Mittag--Leffler expansion \cite{SzK,SzSz2} of the spectral function of
the bubble graph.

\begin{figure}[!ht]
  \centering
  \includegraphics*[32mm,228mm][164mm,279mm]{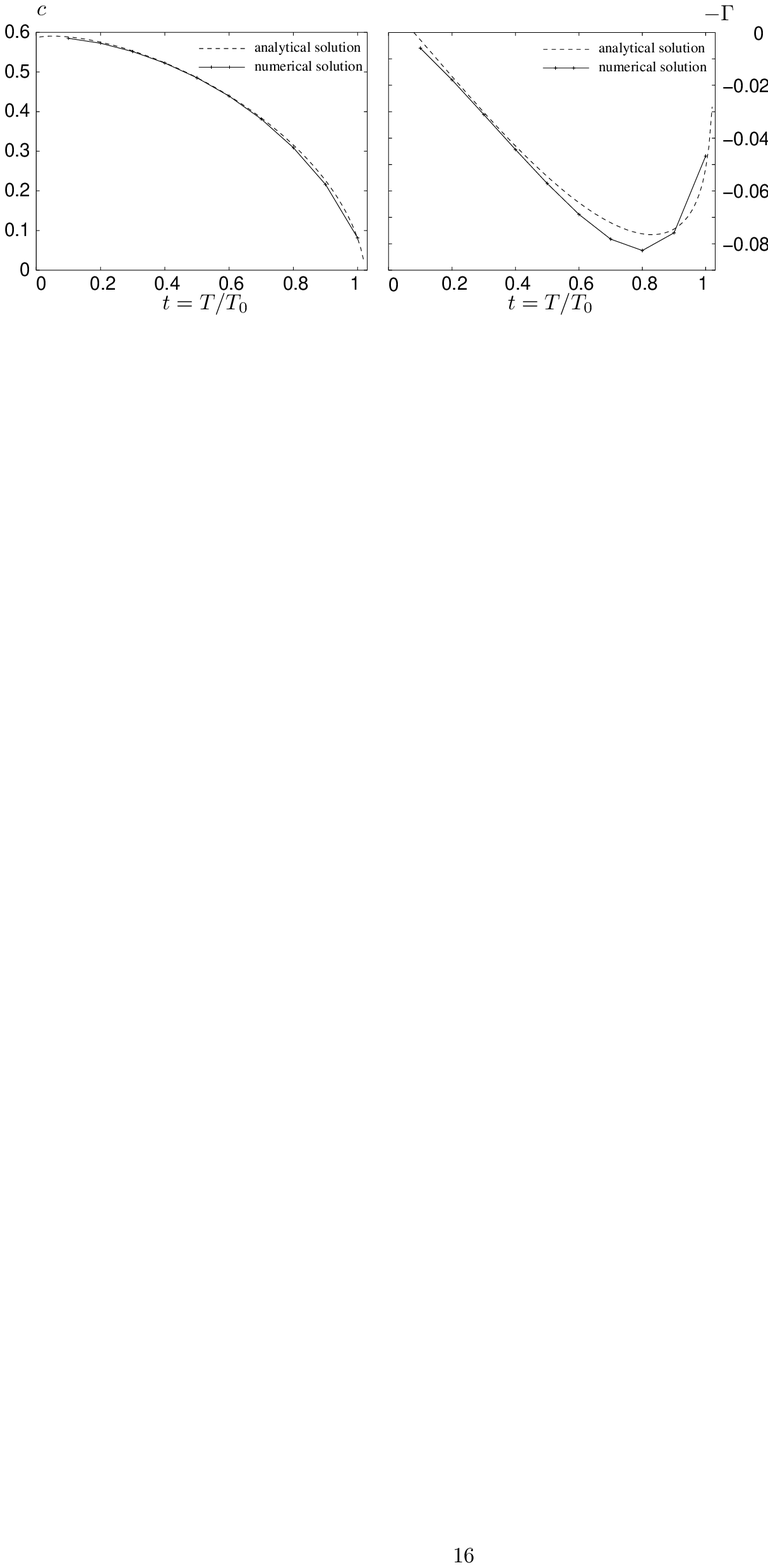}
  \caption{The sound velocity $c$, and the parameter $-\Gamma$ of the
    particle density excitations of the $^{23}\mathrm{Na}$ (polar)
    system in units of $k_B T_0 \lambda_0/\hslash$, and as the
    function of $t=T/T_0$.}
  \label{fig:ceppd}
\end{figure}

The particle density excitations show the same character as their
ferromagnetic counterparts (or as the density excitations in a scalar
gas). Namely they have a linear energy spectrum and linear damping
rate in the small wavenumber limit. The numerical solution and the
analytical approximation from the Mittag--Leffler series
\cite{SzK,SzSz2} is plotted in Fig. \ref{fig:ceppd}, where $c$ and
$\Gamma$ are defined as in Eq. \eqref{eq:dwf}.

\subsubsection{Plus one spin transfer excitations}

The $+1$ spin transfer excitations are spin waves, as shown by Eq.
\eqref{eq:dpwm3}. Using Eqs. \eqref{eq:pse1}, \eqref{eqs:dlde} and
\eqref{eq:pchp}, the dimensionless proper self-energy of spin-transfer
$1$ can be brought to the form:
\begin{flalign}
  \label{eq:dse+p}
    \gmat[+][\widetilde\sigma]_{11}(k)&=\beta N_0\big[
    \Theta^{00}_{--}(0,k,0)-c_n\big]
    +(1-\delta)(C_+ - C_0),\\
    \gmat[+][\widetilde\sigma]_{1,-1}(k)&=\beta N_0\big[
    \Theta^{+0}_{-0}(0,k,-k)-c_s\big],\\
    \gmat[+][\widetilde\sigma]_{-1,1}(k)&=\beta N_0\big[
    \Theta^{0+}_{0-}(-k,0,k)-c_s\big],\\
    \gmat[+][\widetilde\sigma]_{-1,-1}(k)&=\beta N_0\big[
    \Theta^{++}_{00}(-k,0,0)-c_n\big]+(1-\delta)(C_+ - C_0).
\end{flalign}
The zeroth order contribution for the anomalous vertex vectors Eq.
\eqref{eq:vectvert+pol} are
\begin{equation}
  \label{eq:vv+zp}
  \vv[][+][\Lambda]_{(0),1}=\sqrt{N_0}\left(\begin{array}{c}
      0\\1\end{array}\right),\qquad \vv[][+][\Lambda]_{(0),-1}=
  \sqrt{N_0}\left(\begin{array}{c}1\\0\end{array}\right).
\end{equation}
Again with the help of Eqs. \eqref{eq:thetasol},
\eqref{eq:dielstar}, \eqref{eq:regpols}, \eqref{eq:pav1} and
\eqref{eq:pav2} the regular polarization, the anomalous vertex and the
proper self-energy can be given explicitly. With the use of these and
Eqs. \eqref{eq:fsfpb} and \eqref{eq:dielmat} the dielectric function
can be expressed. The spin wave excitations of the polar state are
given by the equation:
\begin{equation}
  \label{eq:de+zp}
  \det\mat[+][\eps](\vec{k},\omega)=0.
\end{equation}

Since the $\Pi^{0+}_{(0)+0}$ and $\Pi^{-0}_{(0)0-}$ bubbles appear in
Eq. \eqref{eq:de+zp}, the analytical solution is also obtained from
the asymptotic series expansion \eqref{eq:spfuncdf} of the spectral
function of the bubble graph.

In the polar case $\widehat\sigma^{++}_{11}=\widehat\sigma^{--}_{11}$,
therefore the self energy difference of the internal lines for the two
bubbles are equal, and as a consequence the damping rate of the spin
wave excitations will be exponentially suppressed but with one
characteristic wavenumber value:
$u_c=|\widehat\sigma^{00}_{11}-\widehat\sigma^{++}_{11}|/2$. The
spectrum of spin wave excitations in the polar case can be written in
the form:
\begin{figure}[!ht]
  \centering
  \includegraphics*[33mm,225mm][99mm,279mm]{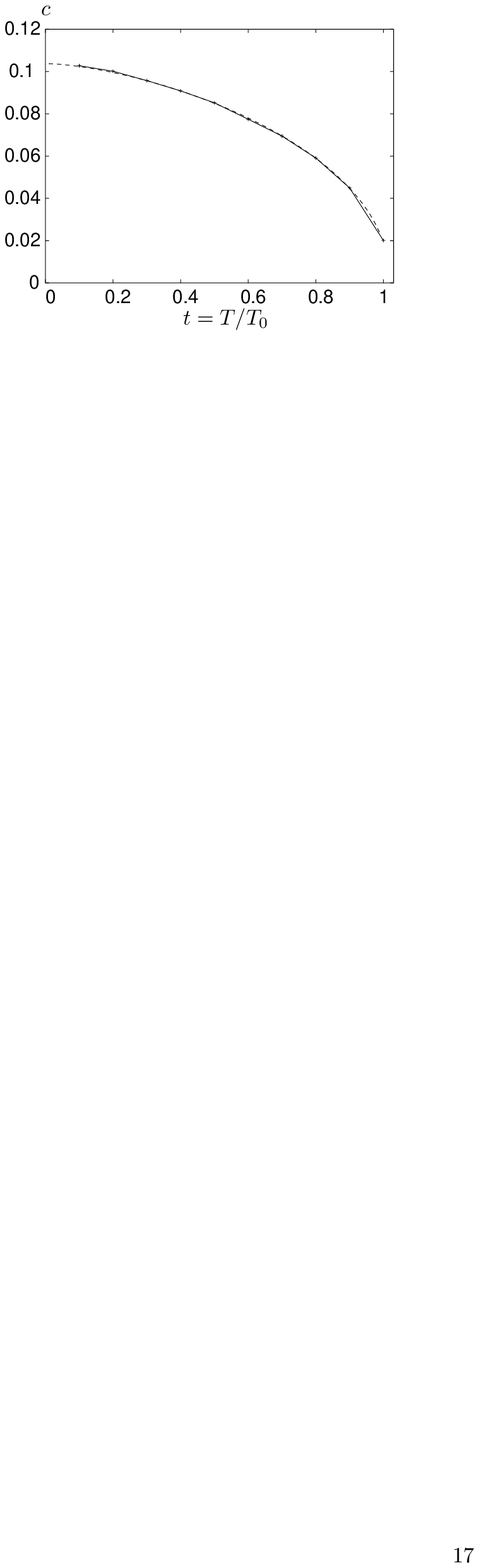}
  \caption{The velocity of the spin-wave excitations of
    the $^{23}\mathrm{Na}$ (polar) system in units of
    $k_B T_0 \lambda_0/\hslash$, and as the function of $t=T/T_0$.}
  \label{fig:swp1}
\end{figure}
\begin{equation}
  \label{eq:swep}
  \omega=c\, |\vec{k}| + i \gamma(|\vec{k}|).
\end{equation}

\begin{figure}[!ht]
  \centering
  \includegraphics*[33mm,223mm][104mm,279mm]{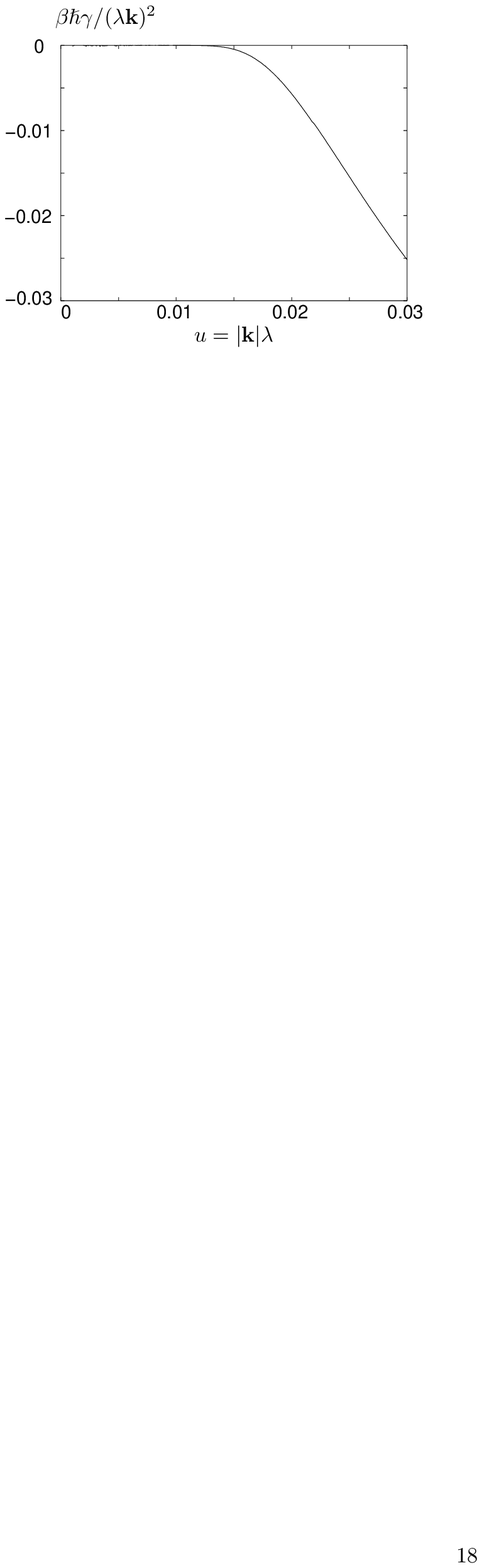}
  \caption{The damping rate of the spin wave excitations at
    temperature $T=0.8\,T_0$ for the $^{23}\mathrm{Na}$ (polar) system
    in units of $(k\lambda)^2/(\hslash \beta)$, and as the function of
    $u=|\vec{k}|\lambda$.}
  \label{fig:swp2}
\end{figure}

The numerical and analytical solutions for the sound velocity of the
spin wave excitations are to be found in Fig. 
\ref{fig:swp1}. The damping
rate is plotted for $T=0.8\,T_0$ in Fig. \ref{fig:swp2}.

\begin{figure}[!ht]
  \centering
  \includegraphics*[30mm,225mm][165mm,279mm]{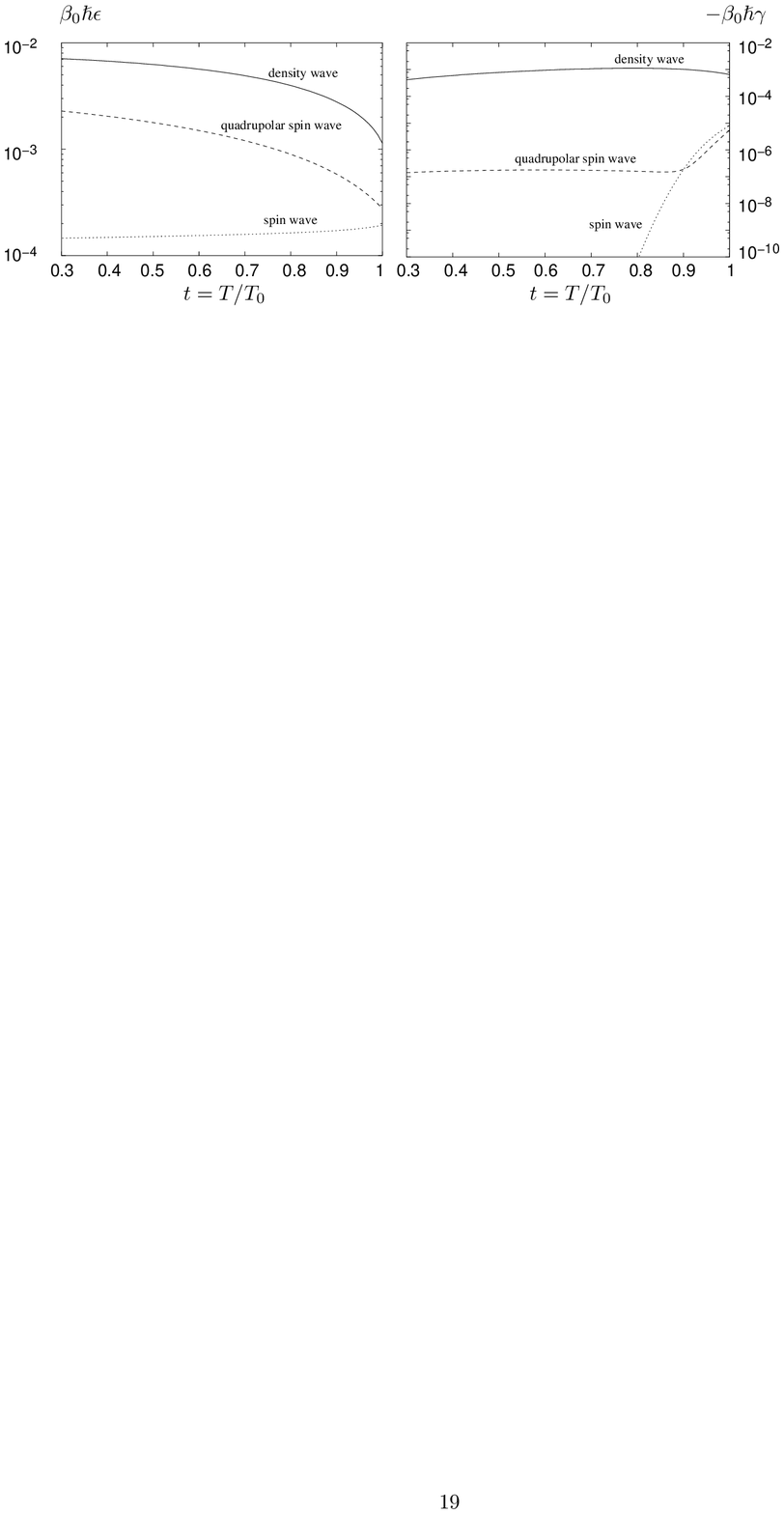}
  \caption{The frequencies ($\epsilon=\Re\,\omega$) and damping rates
  ($\gamma=\Im\,\omega$) as functions of dimensionless temperature
  $t=T/T_0$ of the various collective excitations for
  $k=k_{\mathrm{typ}}=10^{5}\mathrm{m}^{-1}$ of the ferromagnetic
  system. Both quantities are measured in units of $k_B T_0/\hslash$.}
  \label{fig:newfigf}
\end{figure}

\begin{figure}[!ht]
  \centering
  \includegraphics*[30mm,225mm][165mm,279mm]{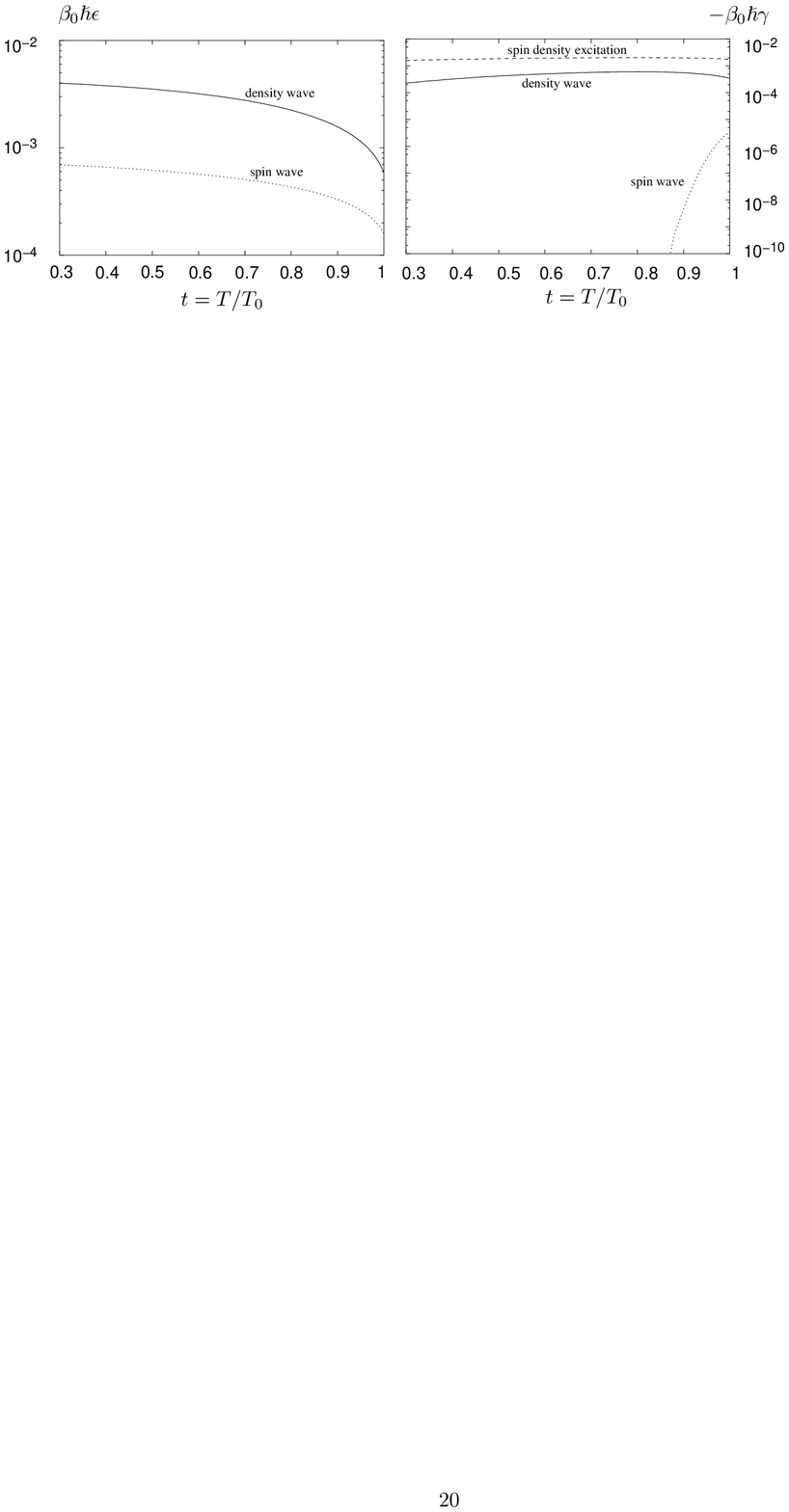}
  \caption{The frequencies ($\epsilon=\Re\,\omega$) and damping rates
  ($\gamma=\Im\,\omega$) as functions of dimensionless temperature
  $t=T/T_0$ of the various collective excitations for
  $k=k_{\mathrm{typ}}=10^{5}\mathrm{m}^{-1}$ of the polar system. Both
  quantities are measured in units of $k_B T_0/\hslash$.}
  \label{fig:newfigp}
\end{figure}

As a summary we compare the frequencies and damping rates of the
various excitations of both the ferromagnetic and polar cases for a
given momentum value, namely
$k=k_{\mathrm{typ}}=10^5\mathrm{m}^{-1}$. In Fig. \ref{fig:newfigf}.
the excitation frequencies and damping rates of the collective
excitations of the ferromagnetic system are plotted. It can be seen,
that for this wavenumber value, density excitations (density
fluctuations) cost much more energy and are relatively short-lived
compared to spin waves and quadrupolar spin waves. These latter
excitations correspond to transverse spin fluctuations of the system
with constant density. In Fig. \ref{fig:newfigp}. the frequencies and
damping rates of the excitations of the polar system are
shown. Density fluctuations cost more energy than transverse spin
fluctuations (in the polar case, we mean transverse as orthogonal to
the spin quantization axis) in this case too and are also shorter
lived than the latter ones. The shortest living excitations in the
polar case however are the non-propagating spin density excitations,
which belong to longitudinal spin fluctuations.

\section{Discussion}
\label{sec:conc}

In this paper we have generalized the self-consistent Hartree-Fock
theory to the symmetry breaking phase of the spin-1 Bose gas. We have
started from the Hartree-Fock equation of state of the ferromagnetic
and polar condensed phases. The Green's functions and correlation
functions have been given, in accordance with the self-consistent
equation of state and with the dielectric formalism, from which the
one-particle and collective excitations of the system have been
determined. The results have been applied to the cases of
${}^{23}\mathrm{Na}$ (polar) and ${}^{87}\mathrm{Rb}$ (ferromagnetic)
gases.

Both in the ferromagnetic and polar cases the behavior of particle
density excitations, belonging to spin transfer $0$, has been found
similar to that of the particle density waves of scalar Bose gases,
where these are the solely existing collective modes. In particular
both their frequency and damping rate depend linearly on wavenumber,
see Eq. \eqref{eq:dwf}, Figs. \ref{fig:cefpd}. and \ref{fig:ceppd}.,
and Ref. \cite{SG,SzK,FRSzG,SzSz2} in the intermediate temperature
region. These collective excitations agree with one-particle
excitations belonging to spin transfer $0$ due to hybridization caused
by the condensate. In the polar case also for spin transfer $0$
another collective mode (independent of particle density waves)
exists, which is overdamped (Fig. \ref{fig:sdip}.) and represents spin
relaxation. This collective mode is related to the pole of the
correlation function \eqref{eq:d2}. Furthermore, it is absent from the
one-particle excitations \cite{SzSz2}, which is the consequence of the
$+\leftrightarrow-$ spin inversion symmetry of the polar phase and is
independent of the approximation. Spin waves are described by the
correlation functions \eqref{eq:d3} and are collective excitations
with spin transfer $1$. They hybridize with the one-particle
excitations with the same spin transfer in both the ferromagnetic and
polar cases. It has been found that with taking into account the
exchange processes the properties of these excitations calculated in
Ref. \cite{SzSz2} essentially change. In the ferromagnetic case,
though the excitation energies continue to show a quadratically
starting energy spectrum, the behavior of the effective mass as a
function of temperature (Fig. \ref{fig:swemf}.) differs qualitatively
from that of the same excitations calculated in the Hartree
approximation, which contains only direct interaction processes. Note,
that in the Bogoliubov approximation the effective mass has been the
bare one \cite{Ho2,OM}. In the polar case the excitation frequencies
depend linearly on wavenumber with a velocity depicted on
Fig. \ref{fig:swp1}. This feature has been found already in the
Bogoliubov and the Hartree approximations \cite{Ho2,OM,SzSz2}. In
these cases, however, the interaction parameter $c_n$ does not show up
in the velocity of the spin wave. It is no more the case in the
present Hartree-Fock approach and the solution drawn on
Fig. \ref{fig:swp1}. includes it. The imaginary part of the spectrum,
i.e. the damping rate is, however, exponentially suppressed
(Fig. \ref{fig:swp2}.), differing qualitatively from the damping rate
in the Hartree approximation, which is linearly dependent on
wavenumber in the relevant temperature region.  The reason of the
observed significant difference is that, while in the Hartree
approximation all three self-energies belonging to internal lines with
different spin projection are zero, in the Hartree-Fock approximation
only the self-energies of the $\widehat\Gr^{++}_{11}$ and
$\widehat\Gr^{--}_{11}$ propagators are equal in the polar case. This
latter coincidence follows from the $+\leftrightarrow-$ symmetry of
the polar state. The lift of the degeneration results in the
appearance of the characteristic wavenumber under which the damping
rate of the excitations are suppressed. (In the ferromagnetic case all
three internal lines have different self-energies, therefore the
damping rate of both the spin wave and quadrupolar spin wave are
suppressed for small momentum values, see Figs. \ref{fig:dqf}. and
\ref{fig:swdf}.)  This can be regarded as a threshold effect. One
expects that in the intermediate temperature region its occurrence is
not restricted to the Hartree-Fock approximation, but it is generally
present though can be less pronounced. The reason is that in this
temperature region scattering processes at wavenumber greater than the
inverse of the Bogoliubov coherence length make the important
contributions and the use of the Hartree-Fock propagator for the
internal lines is justified.  Collective excitations with spin
transfer $2$ (quadrupolar excitations) are propagating only in the
ferromagnetic case. The frequency of these excitations start with a
gap (since these are non Goldstone excitations) and is parabolic (see
Fig. \ref{fig:gemf}.). Its effective mass is close to that of the
ferromagnetic spin waves, the difference is proportional to the
smaller coupling constant $c_s$.  The existence of a gap and the
quadratic dispersion has been found already in the Bogoliubov and
Hartree approximations \cite{Ho2,OM,SzSz2}.

Some remarks concerning hybridization of one-particle and collective
excitations are in order here. We have used this terminology in the
sense as introduced in case of a scalar condensate. Namely, that due
to the presence of the condensate one can see intermediate states
corresponding to collective excitations in the perturbation series of
the one-particle propagators and similarly, one can identify
one-particle intermediate states in the perturbation series of the
correlation functions describing collective modes. As a result the
one-particle and the collective excitations spectra coincide after the
necessary rearrangement of the perturbation series, carried out within
the framework of the so called dielectric formalism (see for a review
Ref. \cite{Griffin}). In case of the systems with a spinor condensate
there are exceptions when such hybridization does not take place
\cite{SzSz2}. One example is the spin density relaxation in the polar
phase described by the correlation function $D_{zz}$ \eqref{eq:d2},
whose perturbation series does not contain one-particle-like
intermediate states (see Eqs. \eqref{eq:sdep},
\eqref{eq:sdwcp}). Another example is provided by the quadrupolar
excitations in the ferromagnetic phase. Though in the perturbation
series $D^Q_{++}$ \eqref{eq:d4} one-particle like intermediate states
occur, it is not true vice versa since the Green's function is proper
in this case (see \ref{sssec:pl2f}. chapter of the Applications). As a
result the spectra of the correlation functions \eqref{eqs:densc} is
more complete than those of the Green's functions. A further advantage
of using the correlation functions is that the physical nature of the
excitation calculated can be seen from the definition of the relevant
correlation function. By these reasons we also have put emphasis on
them in the Section devoted to the applications.

We should also comment on the region of validity of the
self-consistent Hartree-Fock approximation. This approximation can not
be used very close to zero temperature. In this regime other processes
give significant contributions to the self-energies (Beliaev
processes). The Hartree-Fock approximation is expected to provide
quantitatively good results when the mean-field coherence lengths are
the longest length scales in the system. It can be seen in Figs.
\ref{fig:clf}. and \ref{fig:scf}. that this is the case above
$T\approx0.3 T_0$. In this model approximation it is found that the
transition is a first order one for both the ferromagnetic and polar
cases, such as for the scalar gas \cite{FRSzG}. On the other hand the
transition is proved to be second order for the polar case and weakly
first order for the ferromagnetic case in the self-consistent Hartree
approximation \cite{SzSz2}. Since both approximations represent
improved mean-field approaches they are not applicable near the
transition point, where a more sophisticated treatment is needed, which
can also decide upon the order of transition.

\section{Acknowledgement}
\label{sec:ack}

The present work has been partially supported by the Hungarian
Research National Foundation under grant No. OTKA T029552.

\appendix

\section{The self-energies in the Hartree-Fock approximation}
\label{sec:sed}

\begin{figure}[!ht]
  \centering
  \includegraphics*[30mm,135mm][190mm,279mm]{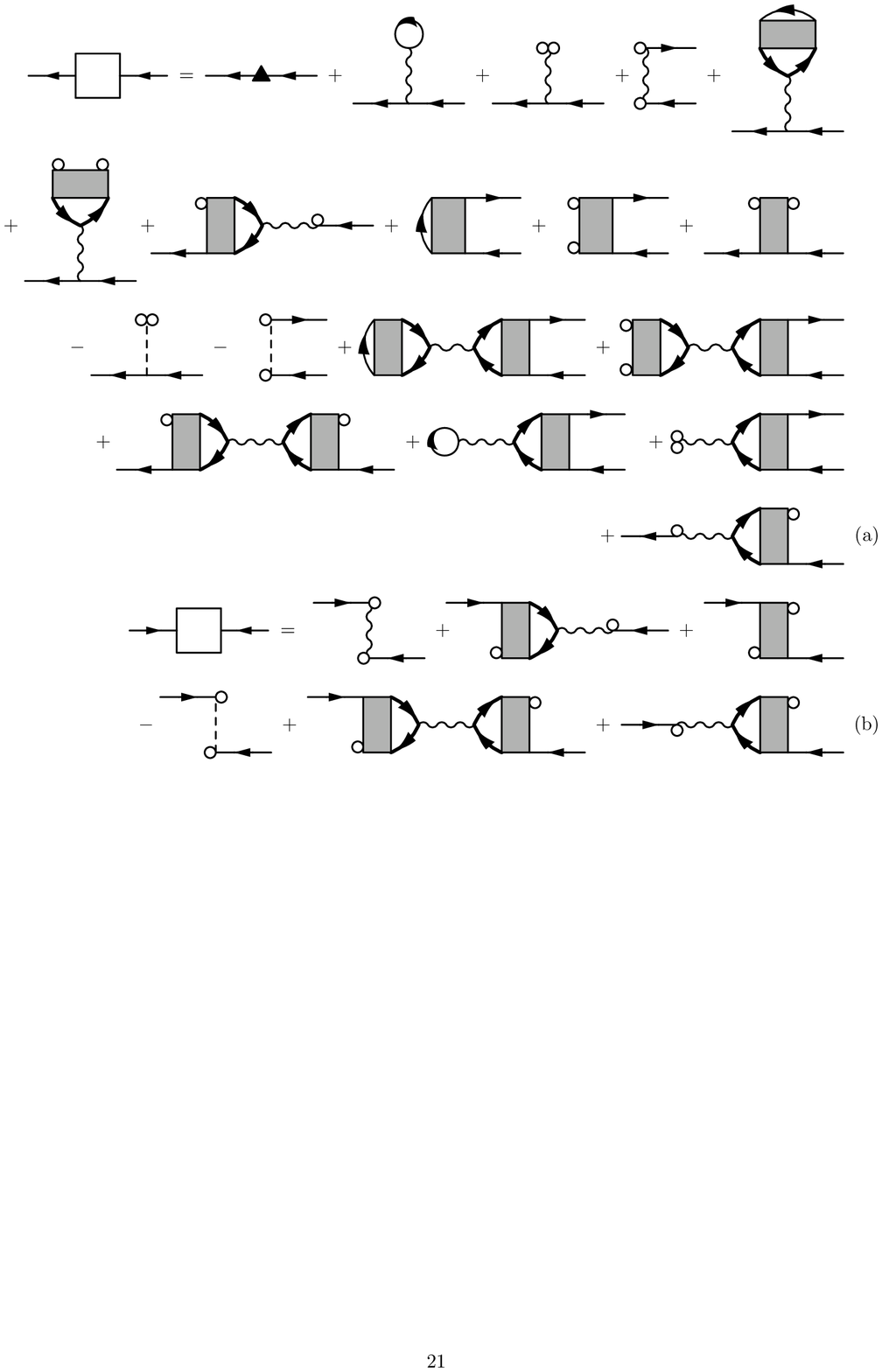}
  \caption{The structure of the self-energies  $\Sigma^{cc}_{11}$ and
    $\Sigma^{cc}_{-1,1}$}
  \label{fig:sed}
\end{figure}

In this appendix we show how the self-energies $\Sigma^{cc}_{11}$ and
$\Sigma^{cc}_{-1,1}$ (with $c=0$ for the polar and $c=+$ for the
ferromagnetic phase), depicted in general in Fig. \ref{fig:nse1}., can
be obtained by changing all condensate circles to external lines, with
the same spin projection, in the tadpole diagrams Fig.
\ref{fig:HFrev}. b). The result of the above process is illustrated in
Fig. \ref{fig:sed}. With the help of the diagramatic equation in Fig.
\ref{fig:HFrev}. a) one can replace the self-energy contribution of
the internal line by the simpler form depicted in Fig.
\ref{fig:intse}. Also using the anomalous vertex vectors, plotted in
Fig. \ref{fig:pav} one arrives at the self-energies as illustrated in
Fig. \ref{fig:nse1}.

\section{The analytical approximation of the bubble graphs with
  different self-energies}
\label{sec:absa}

In this appendix we outline the approximation of the bubble graphs
with different self-energies of the propagators. The contribution of
the analytically continued bubble graph, with internal propagators
carrying spin $s$ and spin $r$, forward and backward, respectively,
can be cast to the form \cite{SzSz2}:
\begin{subequations}
  \label{eqs:bubcont}
  \begin{equation}
    \label{eq:bubcont1}
    c_n \Pi^{sr}_{(0)rs}(u,\Omega)=\frac{\epsilon_n \sqrt{t}}{12 \Gamma(3/2)
      \zeta(3/2) u}\Big[\int_\infty^{y_0}
    R_{\ws^{rr}_{11}}(z)dz-\int_\infty^{y_1}R_{\ws^{ss}_{11}}(z)dz\Big],
  \end{equation}
  with the notations
  \begin{align}
    y_0&=\frac{1}{2 u}[u^2(\Omega-1)-\Delta \ws],\\
    y_1&=\frac{1}{2 u}[u^2(\Omega+1)-\Delta \ws],\quad \text{and}\\
    \Delta \ws &= \ws^{ss}_{11}-\ws^{rr}_{11}.
  \end{align}
\end{subequations}
The  function
\begin{equation}
  \label{eq:spfunc}
  R_{\ws}(z) = \int_{-\infty}^{\infty}\frac{2x}{e^{x^2+\ws}-1}
  \frac{1}{x-z}dx, \quad (\text{for}\quad \Im\;z>0)
\end{equation}
(analytical in the upper half-plane) is to be continued to the whole
complex plane. We are interested in the $\Im\;\Omega<0$ values, since
the elementary excitations have finite lifetimes. Therefore $z$ also has
a negative imaginary part, therefore the 
function reads as:
\begin{subequations}
  \label{eqs:spfd}
  \begin{align}
    R_{\ws}(z)&=R'_{\ws}(z)+R''_{\ws}(z),\quad\text{with}\\
    R'_{\ws}(z) &= \int_{-\infty}^{\infty}\frac{2x}{e^{x^2+\ws}-1}
    \frac{1}{x-z}dx\label{eq:spfd1}\\
    R''_{\ws}(z) &=\frac{4 i \pi z}{e^{z^2+\ws}-1}, \quad
    (\text{for}\quad \Im\;z<0).\label{eq:spfd2}
  \end{align}
\end{subequations}
In the long wavelength limit, i.e. when $2 u \ll |\Delta \ws|$, the upper
limits of the integrals in Eq. \eqref{eq:bubcont1} diverge,
therefore an asymptotic series expansion can be used to represent the
integral Eq. \eqref{eq:spfd1}:
\begin{equation}
  \label{eq:spfuncdf}
    R'_{\ws}(z) = -\frac{2}{z^2}\sum_{n=0}^{\infty}
    \frac{\Gamma(n+3/2)F(n+3/2,\ws)}{z^{2n}}.
\end{equation}
Let us define the quantities $\Pi^{'sr}_{(0)rs}$ and
$\Pi^{''sr}_{(0)rs}$, with Eq. \eqref{eqs:bubcont} by replacing $R$
with $R'$, and $R''$, respectively. One obtains
\begin{equation}
  \label{eq:bubpp}
  c_n\Pi^{''sr}_{(0)rs}(u,\Omega)=\frac{2\pi i\epsilon_n\sqrt{t}}{12\Gamma(3/2)
      \zeta(3/2) u}\ln\frac{1-e^{-y_0^2+\ws^{rr}_{11}}}
    {1-e^{-y_1^2+\ws^{ss}_{11}}}.
\end{equation}
In the same limit, i.e. when $2 u \ll |\Delta \ws|$, one can further
approximate $c_n\Pi^{''sr}_{(0)rs}$ with a result:
\begin{multline}
  \label{eq:bubppf}
   c_n\Pi^{''sr}_{(0)rs}(u,\Omega)=\frac{2\pi i\epsilon_n\sqrt{t}}
   {12\Gamma(3/2)\zeta(3/2) u}e^{-\frac{\Delta\ws^2}{4u^2}}\Bigg[
   \exp\bigg(-\frac{u^2(\Omega-1)^2-2\Delta\ws(\Omega-1)-4\ws^{ss}_{11}}{4}
   \bigg)\\-\exp\bigg(-\frac{u^2(\Omega+1)^2-2\Delta\ws(\Omega+1)-4
     \ws^{rr}_{11}}{4}\bigg)\Bigg].
\end{multline}
Taking the real part of $\Omega$ positive, and with the assumption
that its imaginary part is small one can see that the quantity in the
square brackets is not important for the considerations made in Sec.
\ref{sec:appl}.

We have considered the analytical approximation of the retarded part
of the bubble graph's contribution with differing self-energies in the
forward and backward propagating lines, i.e. with nonzero spin
transfer. The approximation for such bubble graphs with the same
self-energies in their internal lines, i.e. those occurring in zero
spin transfer processes, are obtained by a Mittag-Leffler expansion,
discussed in detail e.g. in Ref. \cite{SzK}.

\end{document}